\documentclass[11pt]{article}

\usepackage{geometry}
\usepackage{booktabs,calc,relsize}

\usepackage[pdftex]{graphicx}

\usepackage{tikz}
\usetikzlibrary{matrix}
\graphicspath{{./images/}}

\usepackage{algorithm}
\usepackage{graphbox}
\usepackage{rotating}

\usepackage{caption}
\usepackage{amssymb,amsmath,amsthm} 
\usepackage{bm}
\usepackage[mathscr]{eucal}
\usepackage{colortbl}
\usepackage{color}
\usepackage{pifont}

\usepackage{epstopdf,hyperref,enumerate,polynom,polynomial}
\usepackage{multirow,minitoc,fancybox,array,multicol}

\usepackage{subcaption}
\usepackage{paralist}

\usepackage{tikz}
\usepackage{pgfplots}
\usepackage{pgfplotstable}
\usepackage{pgfgantt}
\pgfplotsset{compat=newest}

\usetikzlibrary{arrows,shapes,positioning,shapes.geometric}
\usetikzlibrary{decorations.markings}
\usetikzlibrary{shadows,automata}
\usetikzlibrary{patterns}
\usetikzlibrary{trees,mindmap,backgrounds}
\usetikzlibrary{decorations.text}
\usetikzlibrary{%
    decorations.pathreplacing,%
    decorations.pathmorphing%
}
\tikzset{no shadows/.style={general shadow/.style=}}
\usepackage{listings}
\DeclareFixedFont{\ttb}{T1}{txtt}{bx}{n}{8} 
\DeclareFixedFont{\ttm}{T1}{txtt}{m}{n}{8}  

\usepackage{lineno}
\usepackage{color}
\definecolor{deepblue}{rgb}{0,0,0.5}
\definecolor{deepred}{rgb}{0.6,0,0}
\definecolor{deepgreen}{rgb}{0,0.5,0}
\definecolor{darkcandyapplered}{HTML}{A40000}

\newcommand{\lt}{\left}
\newcommand{\rt}{\right}

\newcommand{\fr}{\frac}

\newcommand{\bc}{\begin{compactenum}[\quad--]}
\newcommand{\ec}{\end{compactenum}}

\newcommand{\AI}{{\it Auto Innovative}}
\newcommand{\AS}{{\it Auto Sprawl}}

\usepackage{soul}
\usepackage{catoptions}
\makeatletter
\def\Autoref#1{%
  \begingroup
  \edef\reserved@a{\cpttrimspaces{#1}}%
  \ifcsndefTF{r@#1}{%
    \xaftercsname{\expandafter\testreftype\@fourthoffive}
      {r@\reserved@a}.\\{#1}%
  }{%
    \ref{#1}%
  }%
  \endgroup
}
\def\testreftype#1.#2\\#3{%
  \ifcsndefTF{#1autorefname}{%
    \def\reserved@a##1##2\@nil{%
      \uppercase{\def\ref@name{##1}}%
      \csn@edef{#1autorefname}{\ref@name##2}%
      \autoref{#3}%
    }%
    \reserved@a#1\@nil
  }{%
    \autoref{#3}%
  }%
}
\makeatother

\newcounter{tmp}

\tikzset{D/.style={
preaction={draw=blue,line width=1pt},
preaction={decoration={contour lineto closed, contour distance=6pt},
decorate,
},
postaction={
insert path={%
\pgfextra{%
\pgfinterruptpath
\begin{scope}[opacity=0.5, transparency group]
\path[fill=blue,even odd rule] 
\mySecondList \myList 
;
\end{scope}
\endpgfinterruptpath}
}},
}}

\tikzset{EDR/.style={
preaction={decoration={contour lineto closed, contour distance=6pt},
decorate,
},
postaction={
insert path={%
\pgfextra{%
\pgfinterruptpath
\path[pattern=north west lines, pattern color=gray,opacity=.7,even odd rule] 
\mySecondList \myList 
;
\endpgfinterruptpath}
}},
}}

\makeatletter
\def\pgfdecoratedcontourdistance{0pt}
\pgfset{
  decoration/contour distance/.code=%
    \pgfmathsetlengthmacro\pgfdecoratedcontourdistance{#1}}
\pgfdeclaredecoration{contour lineto closed}{start}{%
  \state{start}[
    next state=draw,
    width=0pt,
    persistent precomputation=\let\pgf@decorate@firstsegmentangle\pgfdecoratedangle]{%
    \pgfextra{\xdef\myList{}\xdef\mySecondList{}}
    \pgfextra{\setcounter{tmp}{0}}
    \pgfpathmoveto{\pgfpointlineattime{.5}
      {\pgfqpoint{0pt}{\pgfdecoratedcontourdistance}}
      {\pgfqpoint{\pgfdecoratedinputsegmentlength}{\pgfdecoratedcontourdistance}}}%
  }%
  \state{draw}[next state=draw, width=\pgfdecoratedinputsegmentlength]{%
    \ifpgf@decorate@is@closepath@%
      \pgfmathsetmacro\pgfdecoratedangletonextinputsegment{%
        -\pgfdecoratedangle+\pgf@decorate@firstsegmentangle}%
    \fi
    \pgfmathsetlengthmacro\pgf@decoration@contour@shorten{%
      -\pgfdecoratedcontourdistance*cot(-\pgfdecoratedangletonextinputsegment/2+90)}%
    \pgfpathlineto
      {\pgfpoint{\pgfdecoratedinputsegmentlength+\pgf@decoration@contour@shorten}
      {\pgfdecoratedcontourdistance}}%
    \stepcounter{tmp}
    \pgfcoordinate{muemmel\thetmp}{\pgfpoint{\pgfdecoratedinputsegmentlength+\pgf@decoration@contour@shorten}
      {\pgfdecoratedcontourdistance}}
    \pgfcoordinate{feep\thetmp}{\pgfpoint{\pgfdecoratedinputsegmentlength}{0pt}}      
    \pgfextra{\xdef\myList{\myList (muemmel\thetmp) -- }%
        \xdef\mySecondList{\mySecondList (feep\thetmp) -- }}
    \ifpgf@decorate@is@closepath@%
      \pgfpathclose
      \pgfextra{\xdef\myList{\myList cycle}%
      \xdef\mySecondList{\mySecondList cycle}}
    \fi
  }%
  \state{final}{\pgfextra{
  }}%
}
\makeatother
\tikzset{
  contour/.style={
    decoration={
      name=contour lineto closed,
      contour distance=#1
    },
    decorate}}

\usepackage{authblk}
\title{Activity-based contact network scaling and epidemic propagation in metropolitan areas}

\author[1]{Nishant Kumar}
\author[2]{Jimi B.\ Oke}
\author[3]{Bat-hen Nahmias-Biran}
\affil[1]{ETH Zurich, Future Resilient Systems, Singapore-ETH Centre, Singapore 138602}
\affil[2]{Department of Civil and Environmental Engineering,  University of Massachusetts, Amherst, MA 01003, United States}
\affil[3]{Department of Civil Engineering, Ariel University, Ariel 40700, Israel}%

\begin{document}
\maketitle

\begin{abstract}
  Given the growth of urbanization and emerging pandemic threats, more
  sophisticated models are required to understand disease propagation
  and investigate the impacts of intervention strategies across
  various city types. We introduce a fully mechanistic, activity-based
  and highly spatio-temporally resolved epidemiological model which
  leverages on person-trajectories obtained from integrated mobility
  demand and supply models in full-scale cities. Simulating COVID-19
  evolution in two full-scale cities with representative synthetic
  populations and mobility patterns, we analyze activity-based contact
  networks. We observe that transit contacts are scale-free in both
  cities, work contacts are Weibull distributed, and shopping or
  leisure contacts are exponentially distributed. We also investigate
  the impact of the transit network, finding that its
  removal dampens disease propagation, while work is also critical to
  post-peak disease spreading. Our framework, validated against
  existing case and mortality data, demonstrates the potential for
  tracking and tracing, along with detailed socio-demographic
  and mobility analyses of epidemic control strategies.
\end{abstract}

\newpage

\section*{Introduction} 
As the COVID-19 pandemic caused by the SARS-CoV-2 virus continues to ravage populations worldwide, there is a
dire need to better understand the propagation of epidemics in order to mitigate its inimical effects in various
communities.  Cities, in particular, have been hit hard due to their population density and extensive mass
transit systems \cite{sahin20202019}.  Governments across the world have responded by imposing
varying levels of mobility restrictions and social distancing \cite{mckibbin2020global}.  Large amounts of data
have been generated to track the spread of the disease (infection, recovery and death rates). Given the mixed
results from these interventions, more sophisticated tools are required to enable decisionmakers to accurately
predict the trajectory of the disease, as well as model the unknown effects of  mitigating strategies.
Hence, agent-based simulations (ABMs) are critical to this effort.
With modern computational advances, ABMs have
demonstrated great potential in accurately tracking the spread of an epidemic at multiple levels \cite{hackl2019epidemic}.
Many of the prior agent-based models were low-resolution and relied on broad
approximations regarding the trajectories of the populations being represented \cite{hackl2019epidemic}.
Recent high-resolution efforts, however, have taken advantage of detailed cellphone data, as well as miscroscopic
traffic simulations \cite{muller2020mobility}. With a  granular representation of agent movements, more
realistic epidemic simulations can be conducted. Modeling epidemic spreading using an agent-based approach
pursues the progression of a disease through each individual, and tracks the contacts of each individual with
others in the relevant social and geographical networks \cite{smieszek2009mechanistic}.

Agent-based models capture the complexity of
human mobility and social patterns more richly than the classical approaches \cite{mckibbin2020global,hackl2019epidemic}.  Mesoscopic and
microscopic transportation models provide an even more detailed representation of the activity behavior and
physical network \cite{adnan2016simmobility}. Only recently have these models been utilized in the study of disease
propagation \cite{hackl2019epidemic}.  While the transportation-based model in \cite{hackl2019epidemic} reasonably approximated the observed event (seasonal influenza in 2016/2017),
socio-demographic factors were not considered with regard to susceptibility. Likewise, the agent interactions
during public transport usage were not considered and only a small scale demonstration was
performed (1\% of the actual population of Zurich).
Most recently, \cite{muller2020mobility} created a contact network
from the trajectories obtained from cellphone data to study the impact of various epidemic mitigation
scenarios related to COVID-19. However, they could not perform detailed contact tracing, and only a sample of the
population was used. Notably, a recent effort modeled the spatial patterns of
the Switzerland 2003/04 influenza epidemic \cite{smieszek2011reconstructing}. Susceptibility varied according to two age groups but contact
duration and intensity were not included in the model. Thus, the time-resolution was only at the day level and
estimates were made regarding contacts by activity (using only a subset of the potential contacts encountered
by each individual). The spread of COVID-19 in Wuhan, China, has also been simulated using an age
and location-specific model \cite{prem2020effect}. However, this effort did not account for the effects of transit and actual person
movements.

\begin{figure}[htp!]
    \centering
   
    \includegraphics[width=0.9\textwidth]{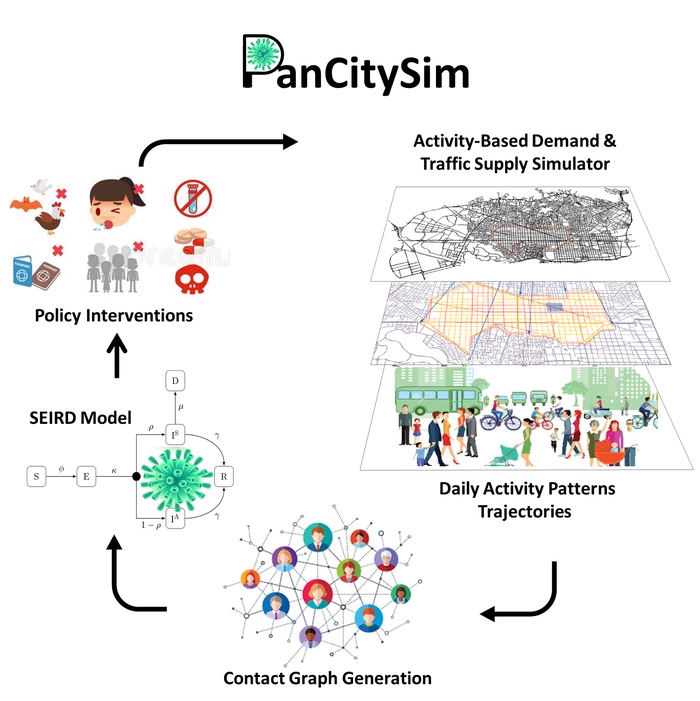}
     \caption{\textbf{PanCitySim framework.} The activity- and agent-based simulator takes as an input land use information, demographic and
economic factors, as well as road and transit network information, and produces daily activity schedules
which are performed on the network to provide multi-modal trajectories for each person.
From the 5-minute person-trajectories, activity-based contact graphs are constructed.
Transmission events are simulated at this resolution.
Other transitions in the SEIRD model are simulated at the end of each day.}
    \label{fig:framework}
 \end{figure}

In this study, we close the gap identified in previous efforts. 
We layer a mechanistic agent-based epidemiological model (PanCitySim) onto highly time-resolved and spatially disaggregated
daily person-trajectories obtained from a microscopic travel demand and
mesoscopic supply simulator, SimMobility \cite{adnan2016simmobility} (see \autoref{fig:framework}).  This
framework provides a realistic representation of person and vehicle movements in an average day for a
full-scale city \cite{oke2019novel}.  We generate activity-specific five-minute contact graphs for the entire population and explore their scaling properties.
We demonstrate how PanCitySim can be used to not only predict
the spatio-temporal dynamics of an epidemic, but also provide detailed contact-tracing and
individual-level analyses of disease impacts.  The epidemiological framework is fully
stochastic, taking into account the latest measurements of relevant parameters, such as the incubation period,
duration of infectiousness, and age-dependent likelihood of symptomaticity and mortality
\cite{muller2020mobility,prem2020effect,verity2020estimates,lauer2020incubation}.
Thus, we exploit the rich activity simulation to represent the activity-based
impacts of  disease transmission.  With PanCitySim, the impacts of a variety of interventions can be evaluated,
e.g. employment-based or age-based restrictions, reduction in mass-transit services, night-time curfews, among
others.  Our framework is  responsive to changes in demand and supply availabilities \cite{nahmias-biran2020evaluating}
and is thus useful for decisionmakers in understanding and mitigating epidemic spreading in metropolitan areas.

 \vspace{1cm}

\section*{Results}
\begin{figure}[htp!]
  \centering
  \begin{subfigure}{.24\linewidth}\centering
    \includegraphics[width=0.6\textwidth,angle=0]{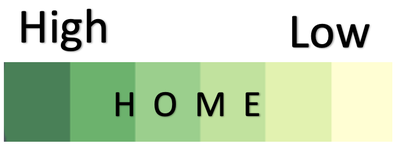}
  \end{subfigure}
  \begin{subfigure}{.24\linewidth}\centering
       \includegraphics[width=0.6\textwidth,angle=0]{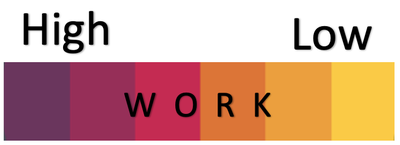}
    \end{subfigure}
  \begin{subfigure}{.24\linewidth}\centering
     \includegraphics[width=.6\textwidth,angle=0]{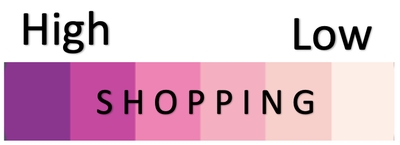}
   \end{subfigure}
   \begin{subfigure}{.24\linewidth}
     \centering
    \includegraphics[width=0.6\textwidth,angle=0]{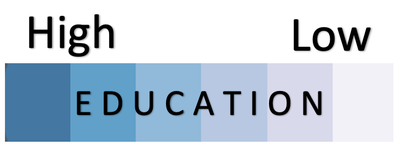}
  \end{subfigure}
  
  \begin{subfigure}{\linewidth}
    \caption{}
    \includegraphics[width=0.24\textwidth]{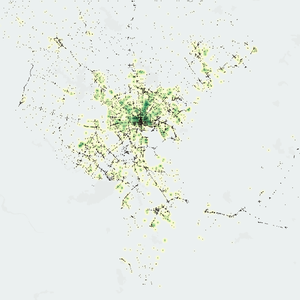}
    \includegraphics[width=0.24\textwidth]{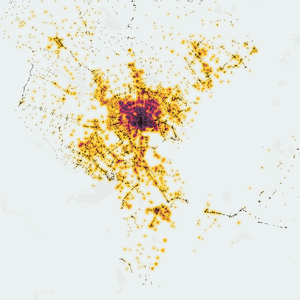}   
    \includegraphics[width=0.24\textwidth]{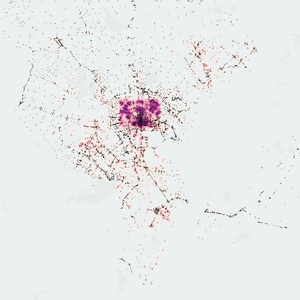}    
    \includegraphics[width=0.24\textwidth]{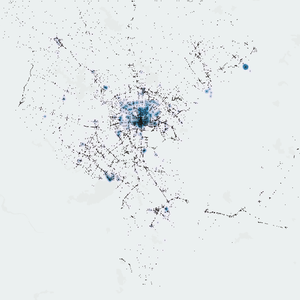}
  \end{subfigure}

  \vspace{1ex}
  
  \begin{subfigure}{\linewidth}
    \caption{}
    \includegraphics[width=0.24\textwidth]{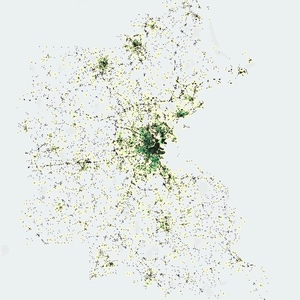}
    \includegraphics[width=0.24\textwidth]{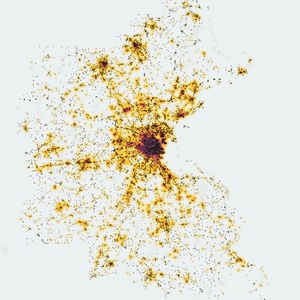}
    \includegraphics[width=0.24\textwidth]{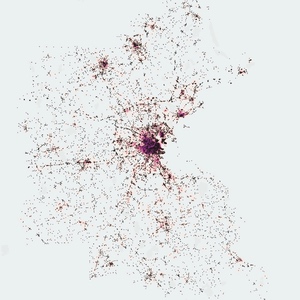}
    \includegraphics[width=0.24\textwidth]{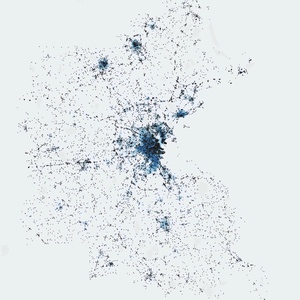}       
  \end{subfigure}

  \vspace{4ex}

  \begin{subfigure}{.45\linewidth}
    \centering \small \textsf{\textit{Auto Sprawl}}
  \end{subfigure}
  \begin{subfigure}{.45\linewidth}
    \centering \small \textsf{\textit{Auto Innovative}}
  \end{subfigure}
  
  \begin{subfigure}{\linewidth}
  \caption{}
    \includegraphics[width=.45\textwidth]{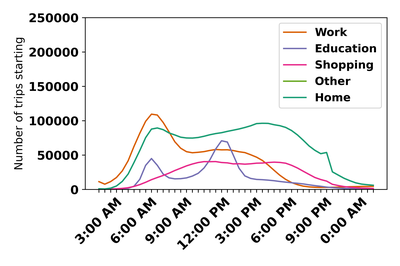}
    \includegraphics[width=.45\textwidth]{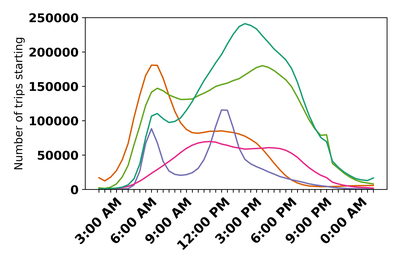}
  \end{subfigure}

  \begin{subfigure}{0.45\linewidth}
  \caption{}
    \includegraphics[width=\textwidth]{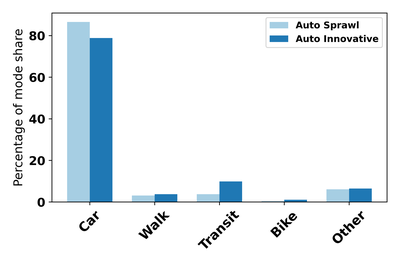}
  \end{subfigure}

  \caption{\textbf{Activity and mobility patterns in prototype cities.}
    Spatial distribution of activity locations in \textbf{a} \textit{Auto Sprawl}, and \textbf{b} \textit{Auto Innovative}.
    The radii of heat maps are the same across both the cities;
    \textbf{c} Number of trips starting at various 30-minute intervals during the day;
    \textbf{d} Mode share for the two cities.
    Daily number of trips in \textit{Auto Sprawl} is $9.72\times10^6$, while in \textit{Auto Innovative} it is $16\times10^{6}$.
    Both cities have a similar trip generation rate of 3.5 per person.
    }

  \label{fig:cities}
\end{figure}

A recent study identified 12 urban typologies based on 69 mobility, socio-demographic, environmental and network structure obtained from over 300 cities \cite{oke2019novel}.
The two auto-dependent typologies consisting of cities largely in the US/Canada were used as the test beds for this study: \AS{} (86\% car mode share)
and \AI{} (78\% car mode share).
\AS{} (e.g.\ Baltimore, Tampa, Raleigh) typifies the lower-density US/Canada cities with low transit usage ($\sim$4\%), while \AI{} (e.g.\ Washington D.C., Atlanta, Boston) are denser cities with an average of 11\% transit mode share.
Prototype cities representing the population, land-use and mobility demand and supply outcomes in both typologies were synthesized
(see Methods and \cite{oke2020evaluating}).
Both prototype cities are built on actual road and transit networks, population microdata and land use categories from representative
(or archetype) cities close to the centroid of their respective typologies.
For \AS, the archetype chosen is the Baltimore Metropolitan Area (population $2.77\times10^{6}$, density $4.11\times10^{2}$km$^{-2}$), while for \AI, it is Greater Boston  (population $4.6\times10^{6}$, density $5.09\times10^{2}$km$^{-2}$).
However, the demand and supply models for both prototype cities were calibrated to fit average typology values, in order to ensure representativeness of overall mobility outcomes.
The spatio-temporal activity and mobility patterns of each city are shown in \autoref{fig:cities}.

\subsection*{Analysis of contact network structure}
\begin{figure}[htp!]
  \centering
  \begin{subfigure}{.45\linewidth}
    \centering \small \textsf{\textit{Auto Sprawl}}
  \end{subfigure}
  \begin{subfigure}{.45\linewidth}
    \centering \small \textsf{\textit{Auto Innovative}}
  \end{subfigure}

   \begin{subfigure}{\linewidth}
    \centering
    \caption{}
    \includegraphics[width=.40\textwidth]{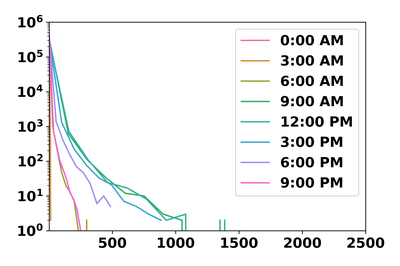} 
    \hspace{1cm}
    \includegraphics[width=.37\textwidth]{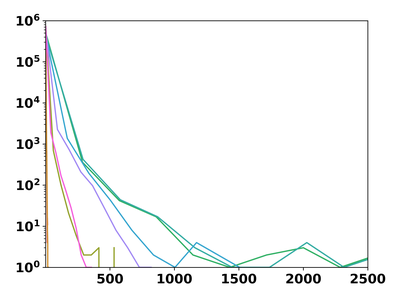}
  \end{subfigure}
  
  \begin{subfigure}{\linewidth}
    \centering
    \caption{}
    \includegraphics[trim={0 0cm 0 0},clip,width=.45\textwidth]{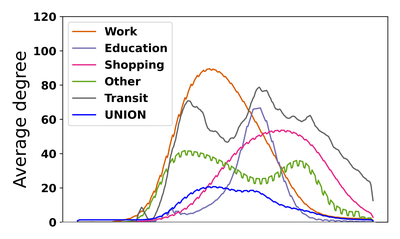}
    \includegraphics[trim={0 0cm 0 0},clip,width=.45\textwidth]{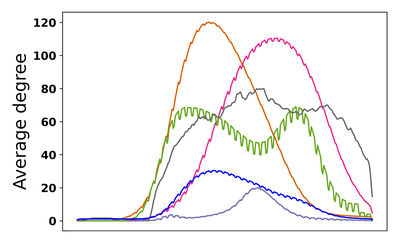}
  \end{subfigure}
    
 \begin{subfigure}{\linewidth}
    \centering
    \caption{}
    \includegraphics[trim={0 0cm 0 0},clip,width=.45\textwidth]{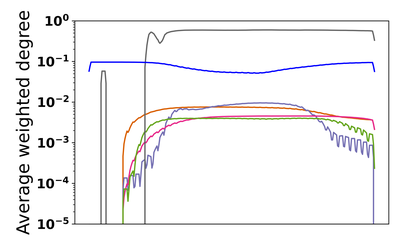}
    \includegraphics[trim={0 0cm 0 0},clip,width=.45\textwidth]{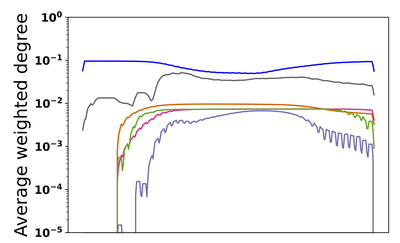}
  \end{subfigure}
  
  \begin{subfigure}{\linewidth}
  \centering
    \caption{}
    \includegraphics[width=.45\textwidth]{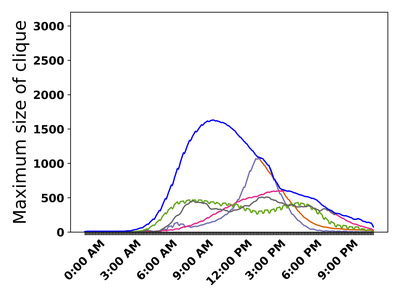}
    \includegraphics[width=.45\textwidth]{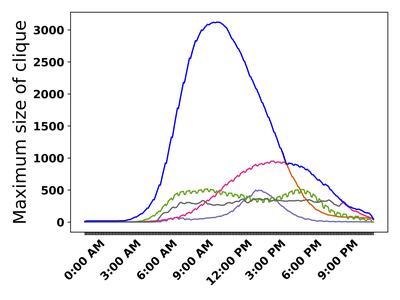}        
  \end{subfigure}
  
  \caption{\textbf{Contact network structure.}
    \textbf{a} Degree distributions of the \textit{Union} contact network for selected times of day;
    \textbf{b} The average degree $\langle k\rangle$ for both cities are shown over the course of the day;
    \textbf{c} The average weighted degree $\langle k_{w} \rangle$ for both cities are shown over the course of the day using log scale). Contacts are weighted by average inverse distances between persons at each node;
    \textbf{d} Diameter (maximum clique size).
    We see that \textit{Work} accounts for the greatest number of contacts per person, particularly during the middle of the day.
    However, the distribution of $\langle k_{w} \rangle$ indicates that \textit{Transit}, more than any other activity, accounts for the closest of contacts.
    For clarity,  plots are smoothed using a moving average of five time steps.
  }
  \label{fig:contact}
\end{figure}

We generate an activity-based contact network for the population in each city every 5 minutes (see Methods).
The \textit{Union} contact network comprises all activities for each individual: \textit{Work}, \textit{Education}, \textit{Shopping}, \textit{Other} and \textit{Transit}. In this paper, the \textit{Transit} activity comprises waiting and the duration of time spent on a bus or trrain.
The spatial resolution of each contact is at the node level (activity location) or in transit vehicle (bus/train) partitions.
\autoref{fig:contact}a shows the degree distribution for the  \textit{Union}  contact network at selected times in both cities.
Further, we plot the average degree $\langle k \rangle$ for each activity contact network, including the \textit{Union}.
We observe that \textit{Work} is responsible for the maximum average number of contacts, peaking at 10 AM with 90 contacts  in
\AS{} and 120  in \AI.
In \AS{}, the activity responsible for the second highest average number of contacts per time step is \textit{Transit}
with 70 contacts on average at morning peak (8 AM) and almost 80 contacts on average at evening peak (3 PM).
In  \AI, however, the activity responsible for second greatest peak contact average is \textit{Shopping}
with 110 contacts per person between 4 and 5 PM.

We also plot the average weighted degree distribution  $\langle k_{w} \rangle$ (\autoref{fig:contact}c), weighting each edge by $\fr 1{d_{nm}}$, where $d_{nm}$ is the estimated distance between persons $n$ and $m$ at a given node at a time step.
The distribution of $\langle k_{w} \rangle$ reveals that \textit{Transit} is responsible for the closest contacts, and that \textit{Shopping} is less important than \textit{Work} in terms of proximity.
The difference between the average separation of contacts in \AS{} is close to two orders of magnitude.
The disparity is less stark (less than one order of magnitude) in \AI{}, given its greater population density).
The maximum number of people who are in contact at any given time of the day is represented by the maximum clique size of the contact graph as shown in \autoref{fig:contact}d.
On average, the maximum number of contacts created in \AS{} is  more than 1600 between 10 and 11 AM, while for \AI{} the maximum number of contacts is twice as high during the same time.

\subsection*{Scaling properties of time-dependent contact graphs}
\begin{figure}[htp!]
  \centering
    \begin{subfigure}{\linewidth}
    \caption{}
    \includegraphics[align=c,width=.3\textwidth]{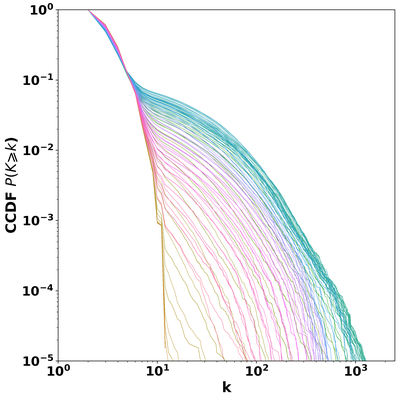}
    \includegraphics[align=c,width=.3\textwidth]{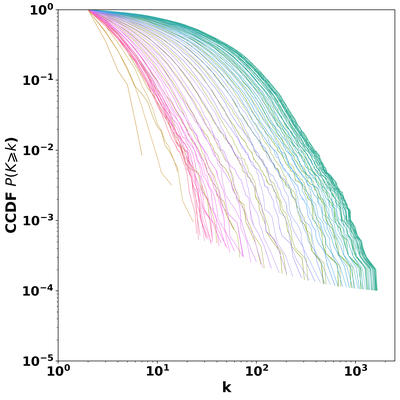}        
    \includegraphics[align=c,width=.3\textwidth]{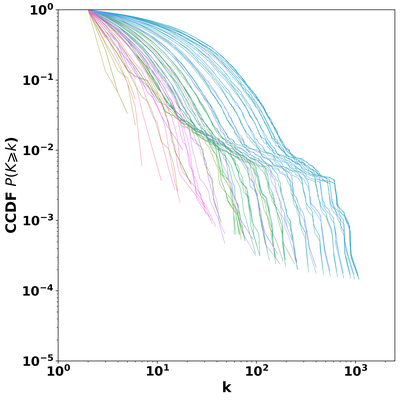}
    \includegraphics[align=c,width=.07\textwidth]{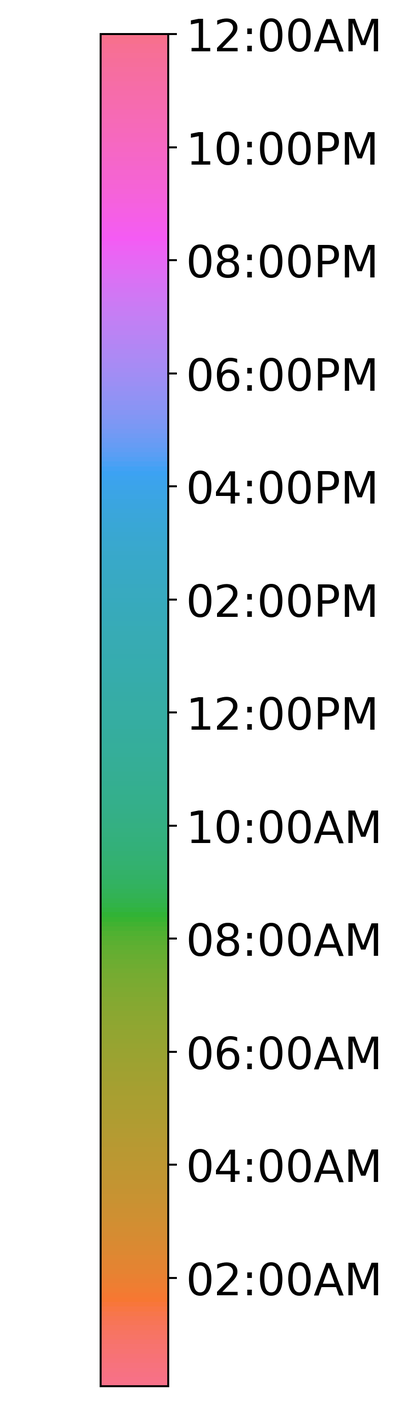} 

     \includegraphics[width=.3\textwidth]{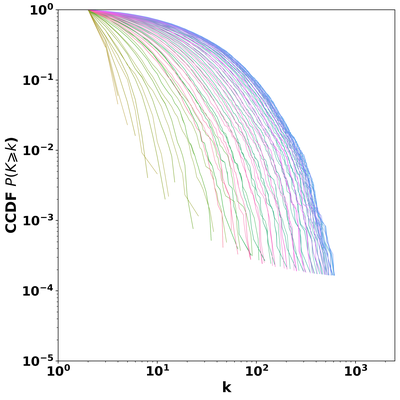}        
    \includegraphics[width=.3\textwidth]{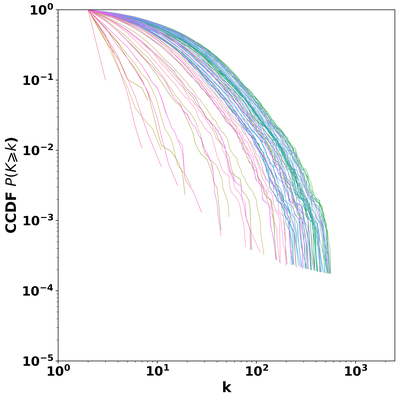}
    \includegraphics[width=.3\textwidth]{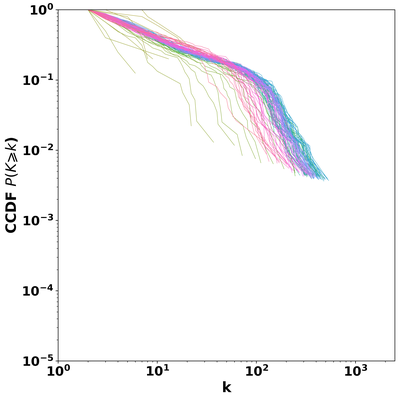}   
  \end{subfigure}

    \begin{subfigure}{\linewidth}
    \caption{}
    \includegraphics[align=c,width=.3\textwidth]{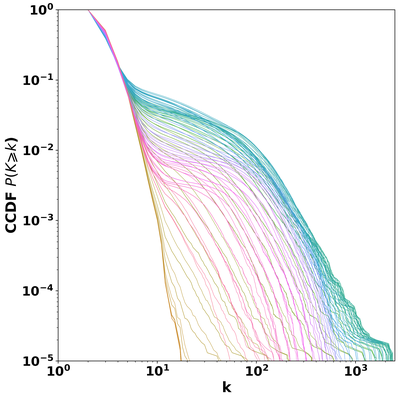}
    \includegraphics[align=c,width=.3\textwidth]{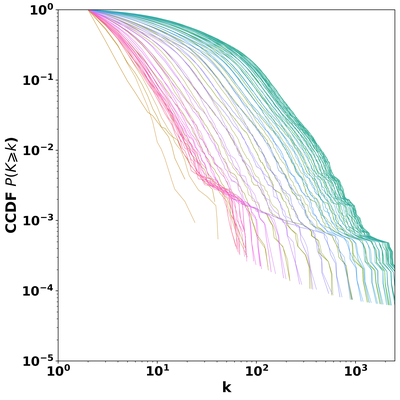}        
    \includegraphics[align=c,width=.3\textwidth]{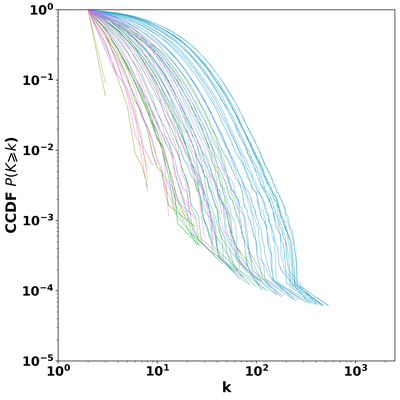}
    \includegraphics[align=c,width=.07\textwidth]{colorbar} 

     \includegraphics[width=.3\textwidth]{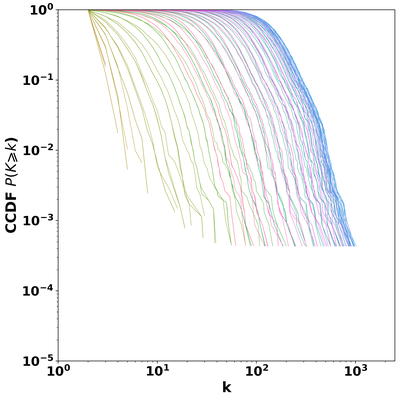}        
    \includegraphics[width=.3\textwidth]{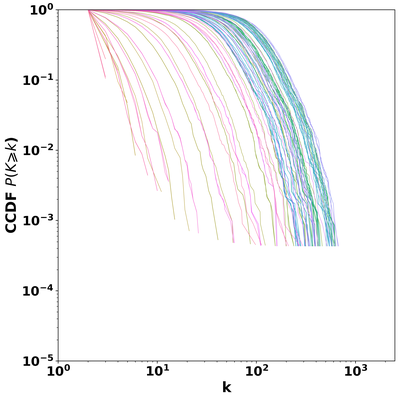}
    \includegraphics[width=.3\textwidth]{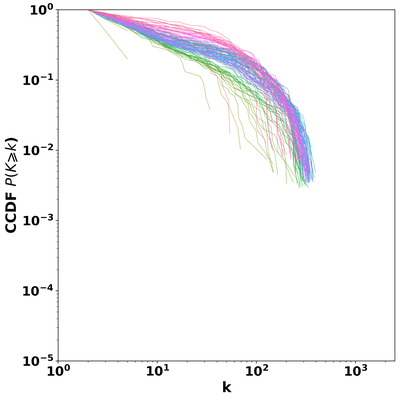}   
  \end{subfigure}

  \caption{\textbf{Activity-based contact networks.} Complementary cumulative distribution (CCDF) plots of time-dependent contacts (clockwise):
    \textit{Union}, \textit{Work}, \textit{Education}, \textit{Transit}, \textit{Other} and \textit{Shopping}.
    \textbf{a} \textit{Auto Sprawl}; \textbf{b} \textit{Auto Innovative}.
  }
  \label{fig:ccdf}
\end{figure}

In order to gain further insights into the contact network structure, we plot the complementary cumulative distribution functions (CCDF) for each time step (5 minutes)
in both cities (\autoref{fig:ccdf}).
The CCDFs are shown for the \textit{Union} contact graphs, as well as for each of the activities including \textit{Transit}.
There are no clear patterns in the union contact graphs, but we hypothesize that the \textit{Work} contact network  follows a Weibull distribution
$p_{k } \propto e^{\lt(-\lambda k \rt)^{\beta}}$,
while the \textit{Transit} contact network obeys the power law,  $p(k) \propto k^{-\alpha}$.
Contact networks for \textit{Shopping} and \textit{Other} activities appear to follow an exponential distribution $p_{k} \propto e^{-\lambda k}$.
In order to test these null hypotheses, we conduct Kolmogorov-Smirnov (KS) tests to determine if alternative distributions provide better fits \cite{alstott2014powerlaw}. 
The KS tests produce p-values based on loglikehood ratios between the alternative and null fits.
In all cases, there was no sufficient evidence to reject the null hypothesis.
The fitted distributions are assumed to be valid from a minimum degree $\hat{k}_{\min}$, which is
determined by minimizing the KS-distance between observed and predicted values based on the power law fit.
While this $\hat{k}_{\min}$ is biased, it facilitates a consistent comparison between candidate distributions \cite{broido2019scalefree}.
Other alternative distributions considered were the lognormal and truncated power law.

\begin{figure}[htp!]
  \centering

  \begin{subfigure}{\linewidth}
    \caption{}
    \includegraphics[align=c,width=.225\textwidth]{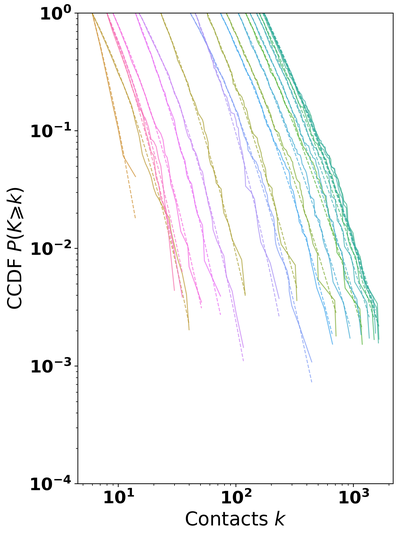}
    \includegraphics[align=c,width=.225\textwidth]{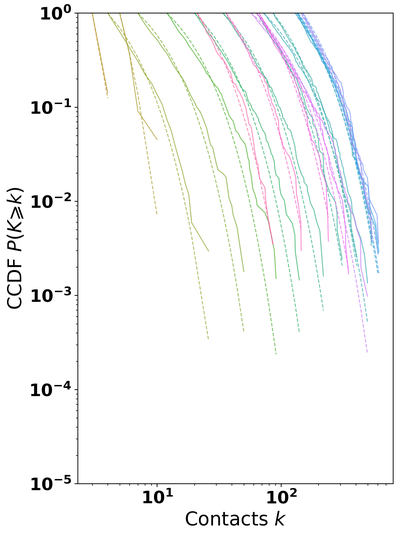}
    \includegraphics[align=c,width=.225\textwidth]{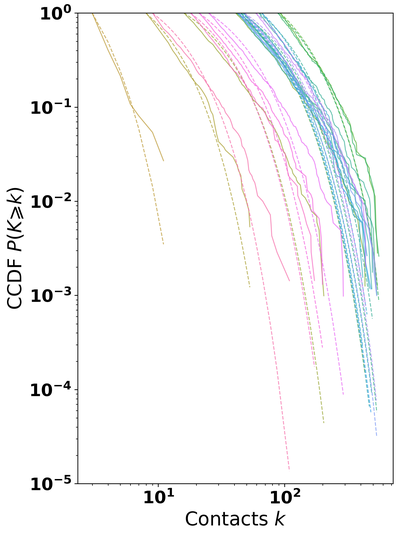}
    \includegraphics[align=c,width=.225\textwidth]{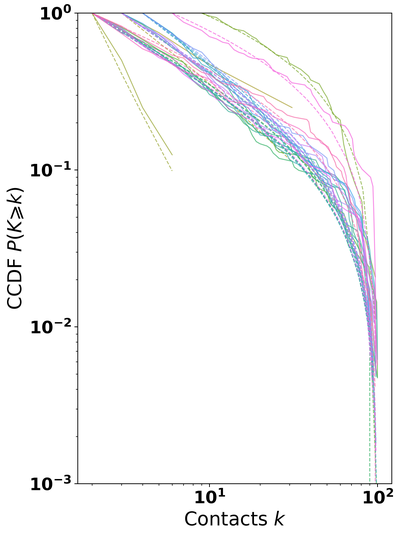}
    \includegraphics[align=c,width=.07\textwidth]{colorbar}
  \end{subfigure}

  \begin{subfigure}{\linewidth}
    \caption{}
    \includegraphics[align=c,width=.225\textwidth]{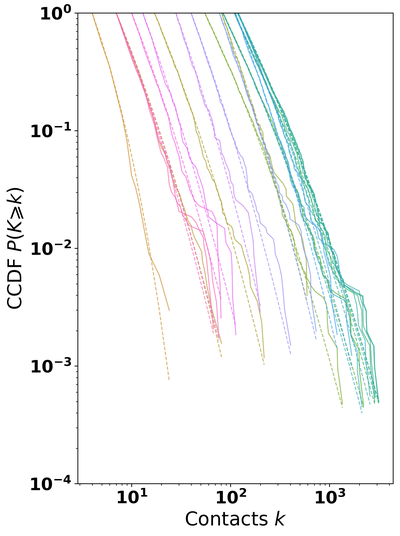}
    \includegraphics[align=c,width=.225\textwidth]{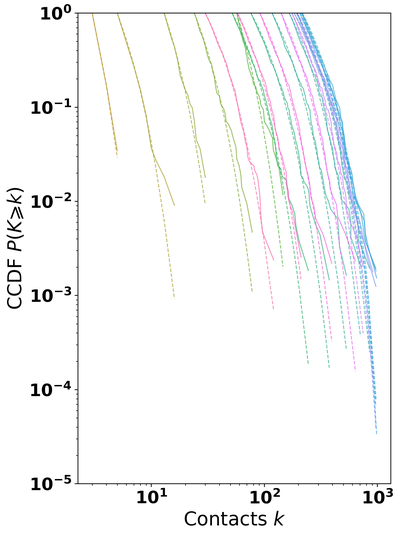}        
    \includegraphics[align=c,width=.225\textwidth]{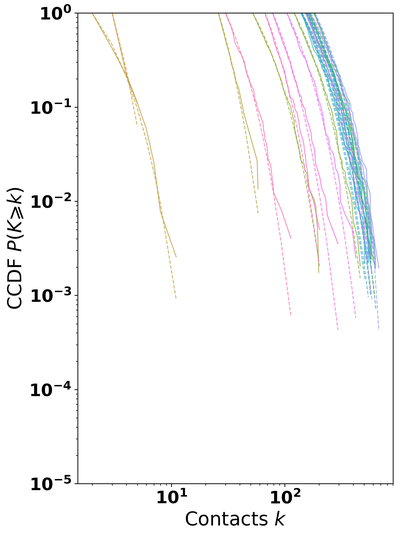}
    \includegraphics[align=c,width=.225\textwidth]{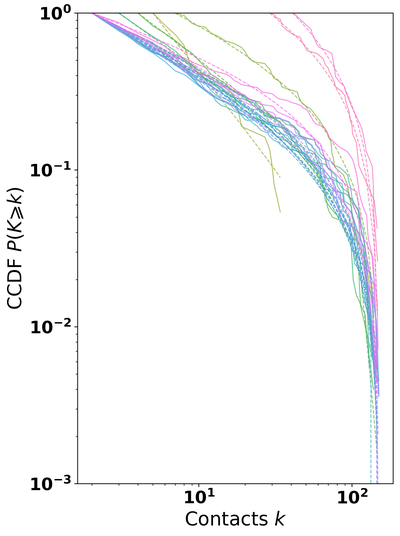}
    \includegraphics[align=c,width=.07\textwidth]{colorbar} 
  \end{subfigure}

  \begin{subfigure}{1.0\linewidth}
    \caption{}
    \includegraphics[align=c,width=.225\textwidth]{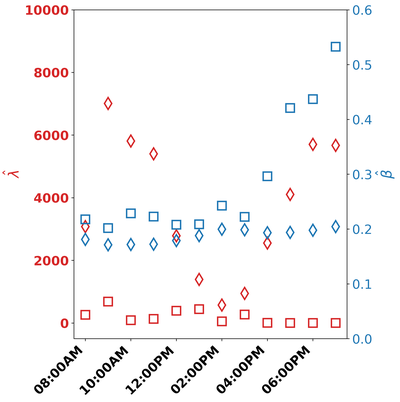}
    \includegraphics[align=c,width=.225\textwidth]{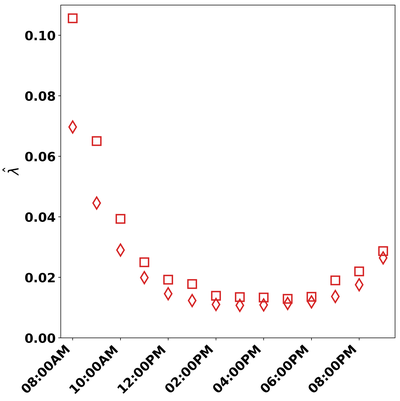}
    \includegraphics[align=c,width=.225\textwidth]{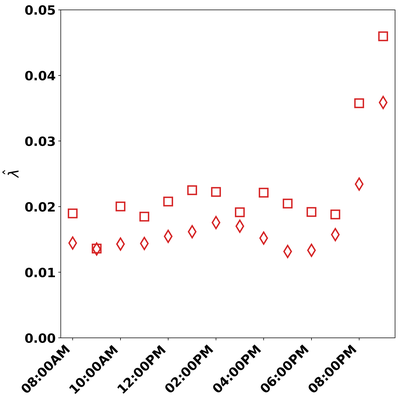}
    \includegraphics[align=c,width=.225\textwidth]{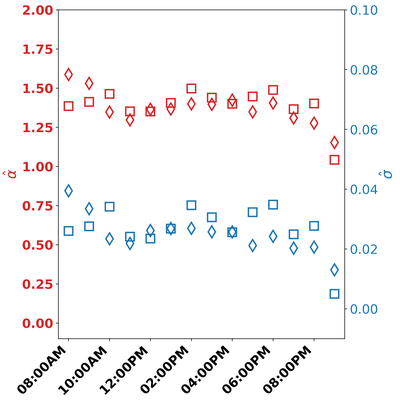}
  \end{subfigure}

  \begin{subfigure}{1.0\linewidth}
    \caption{}
    \includegraphics[align=c,width=.225\textwidth]{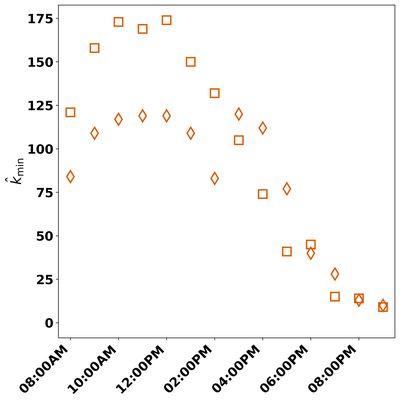}
    \includegraphics[align=c,width=.225\textwidth]{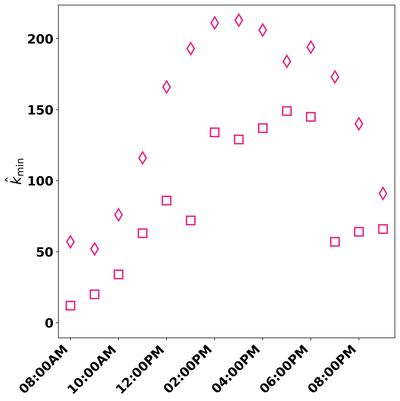}
    \includegraphics[align=c,width=.225\textwidth]{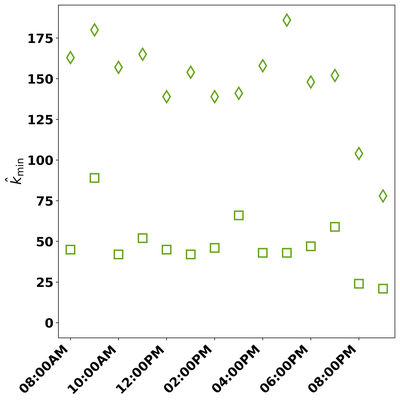}
    \includegraphics[align=c,width=.225\textwidth]{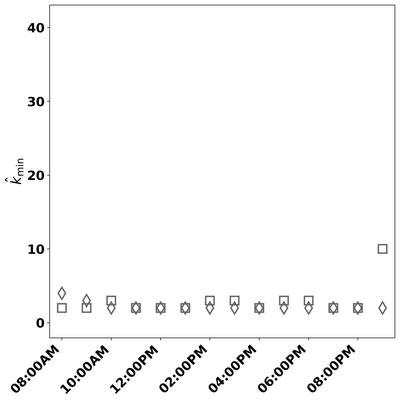}
  \end{subfigure}
  
  \caption{\textbf{Contact network scaling.} Fitted contact network degree distributions (dashed lines) compared to observed distributions from simulation (solid lines) shown
    for the respective activities: \textit{Work}, \textit{Shopping}, \textit{Other} and \textit{Transit}.
    Only hourly fits are displayed for clarity.
    \textsf{\bfseries a } \textit{Auto Sprawl};
    \textsf{\bfseries b } \textit{Auto Innovative}.
    \textsf{\bfseries c } Fitted parameters for  \textit{Auto Sprawl} (square markers $\square$) and \textit{Auto Innovative} (diamond markers  $\Diamond$).
        Work follows a \textit{Weibull} distribution, while \textit{Shopping} and \textit{Other} are fitted to exponential distributions.
        \textit{Transit} is fitted to a power law.
        Parameters are only significant within time periods shown.
        \textsf{\bfseries d } Fitted $\hat{k}_{\min}$ for each activity (\textit{Work}, \textit{Shopping}, \textit{Other} and \textit{Transit}).
        Time-dependence is exhibited for all activities except \textit{Transit}.
  }
  \label{fig:scale}
\end{figure}

The scale-free \textit{Transit} contact graphs are fitted to $p_{k} \propto k^{-\hat\alpha}$, with a cut-off at $k=150$.
Between 8 AM and 8 PM, $\hat\alpha \approx 1.4$ in both cities. 
The  parameter $\hat{k}_{\min}$ is largely time-invariant, averaging 2.4 in \AS{} and 2.2 in \AI{} (\autoref{fig:scale}d).
This indicates that fitted scaling patterns take effect with a minimum of two or three contacts.
For \textit{Shopping}, which follows an exponential distribution, $\hat\lambda$ is similar for both cities at the same time periods.
In the case of \textit{Other}, however, while the trends in \AS{} and \AI{} are similar, $\hat\lambda$ tends to be higher in \AS.
The fitted Weibull parameters for \textit{Work} diverge between both cities (\autoref{fig:scale}c).

\subsection*{Spatio-temporal evolution and age-specific impacts}
\begin{figure}[htp!]
  \centering
  \begin{subfigure}{.45\linewidth}
    \centering \small \textsf{\textit{Auto Sprawl}}
  \end{subfigure}
  \begin{subfigure}{.45\linewidth}
    \centering \small \textsf{\textit{Auto Innovative}}
  \end{subfigure}
  
  \begin{subfigure}{\linewidth}
    \centering
    \caption{}
    \includegraphics[width=.42\textwidth]{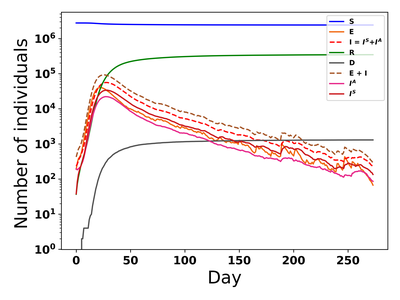}\hfil
    \includegraphics[width=.42\textwidth]{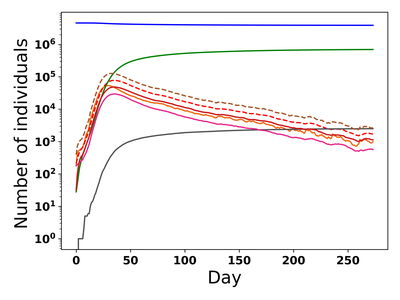}
  \end{subfigure}
  
  \begin{subfigure}{\linewidth}
    \centering
    \caption{}
    \includegraphics[width=.42\textwidth]{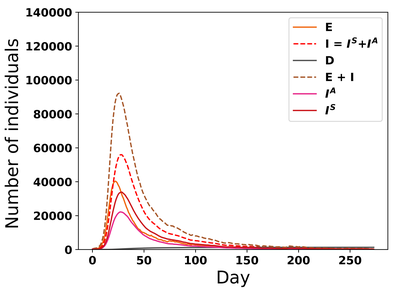}\hfil
    \includegraphics[width=.42\textwidth]{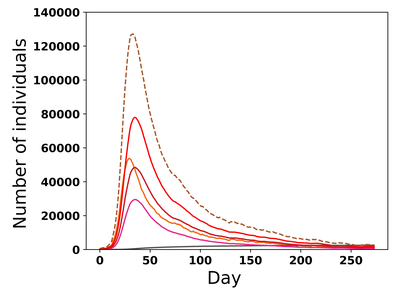}
  \end{subfigure}

  \begin{subfigure}{\linewidth}
    \centering
    \caption{}
    \includegraphics[width=.48\textwidth]{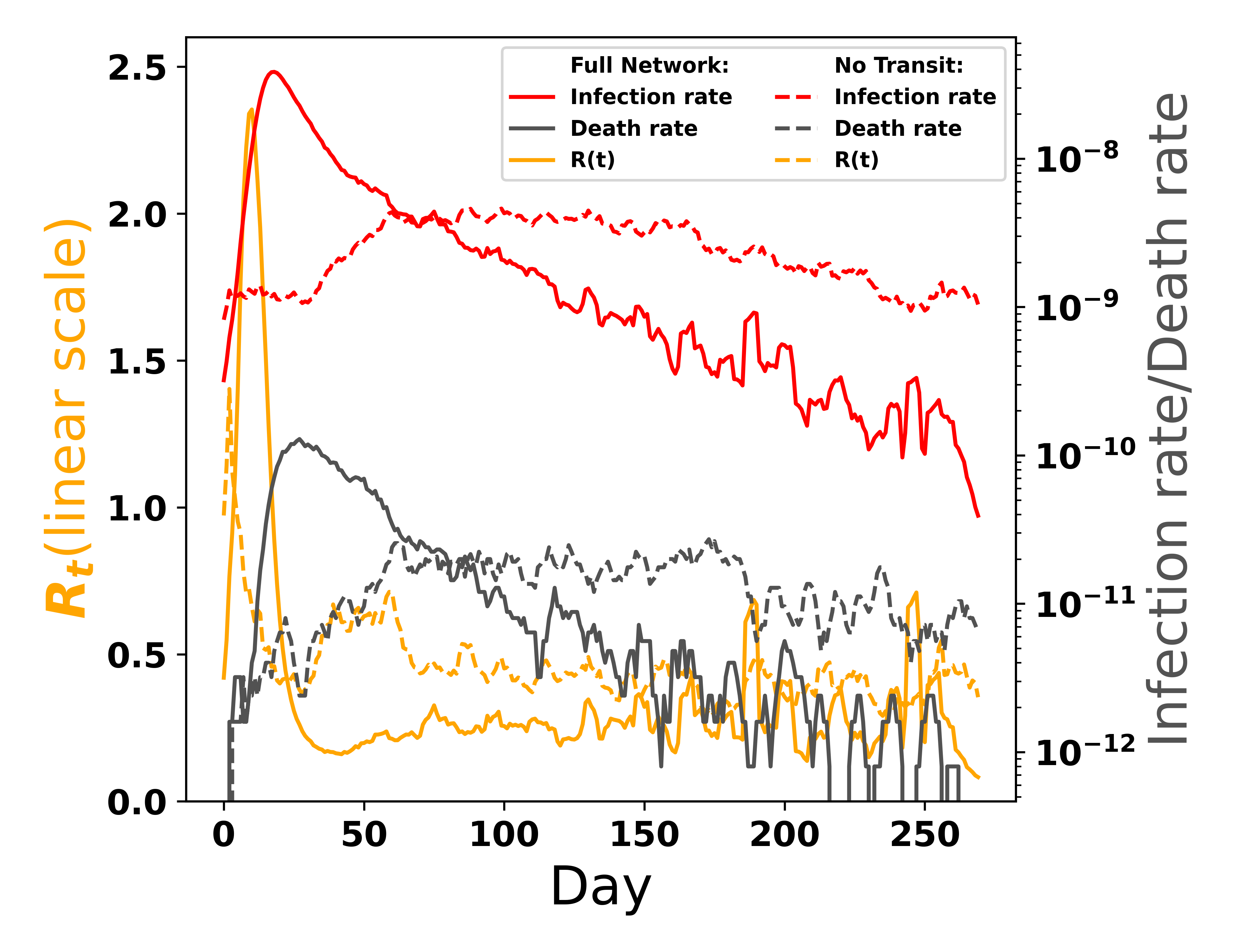}
    \includegraphics[width=.48\textwidth]{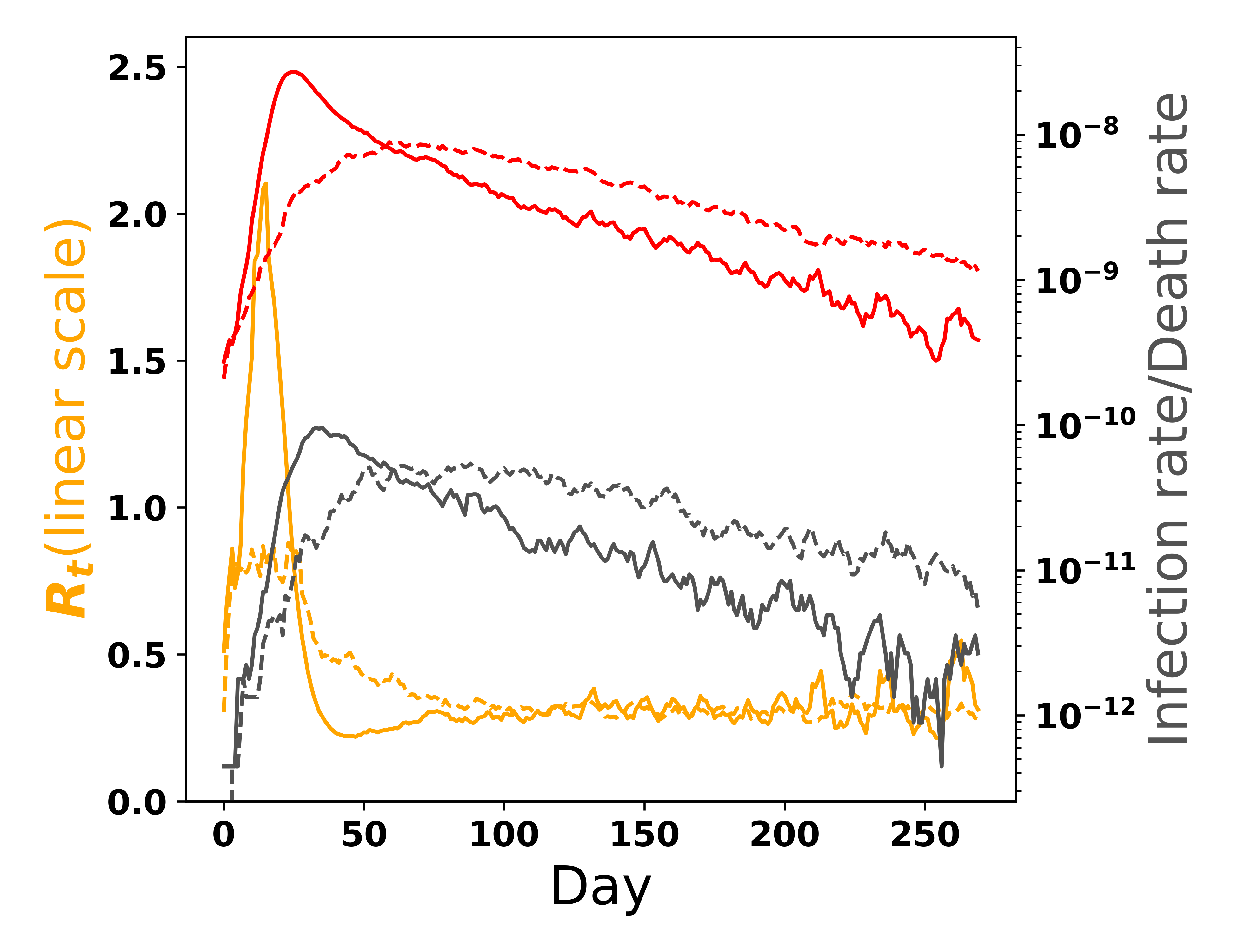}    
  \end{subfigure}
  
  \caption{\small \textbf{Simulated evolution of COVID-19 across two prototype cities.}
    \textbf{a} Log y-axis of population states.
    \textbf{b} Exposed, infected (both asymptomatic and symptomatic cases) and deceased (cumulative).
    \textbf{c} Time-dependent (daily) reproductive rate $R_{t}$, infection and mortality rates.
    The curves are smoothed with a moving average over 5 days.
    The infection rate and death rate are expressed per 100,000 persons.
  }
  \label{fig:temporal}
\end{figure}

We calibrated \AS{} and \AI{} to fit the expected early dynamics of COVID-19 \cite{prem2020effect,kucharski2020early,lin2020explaining,gunzler2020timevarying} (see Methods).
We considered the basic reproductive rate $R_{0}$ (average number of secondary infections per index case), which ranged from
2.29 to 2.57 in \AS, and from 3.03 to 3.35 in \AI.
Given the uncertainty about $R_{0}$, we also considered the time-dependent reproduction number $R_{t}$ (average number of new daily infections per infectious individuals).
We compute the five-day average of $R_{t}$ as a measure of the early propagation of the epidemic, which is  $0.77$ in both cities (see Methods).
Given these starting assumptions, we simulate the epidemic for 270 days in both cities.
The peak number of infections occurs at day 27 in \AS{} and at day 34 in \AI, dissipating after 150 days and 250 days, respectively.
At the peak, there are $9.21\times 10^{4}$ infections (both exposed and infectious individuals) in \AS{} and $1.27\times10^{5}$ infections in \AI{} (\autoref{fig:temporal}b).

We plot the daily infection and mortality rates, along with $R_{t}$ in \autoref{fig:temporal}c.
The infection rate is given by the number of daily new infections as a proportion of the entire population.
The mortality rate is defined as the number of daily deaths also with respect to the entire population.
While both \AS{} and \AI{} have similar early onset dynamics, we observe that disease propagation diverges rapidly between both cities.
In \AS{}, $R_{t}$ peaks at day 9, with an average rate of change of $1.78\times10^{-1}$ day$^{-2}$ from onset to peak.
Post-peak, however, the rate of change is $-9.61\times10^{-3}$ day$^{-2}$.
Furthermore, the slope of the infection rate from onset to peak (day 18) in \AS{} is $5.15\times10^{2}$ day$^{-2}$.
Post-peak, the rate is $4.25\times10^{1}$ day$^{-2}$.
In \AI{}, however, $R_{t}$ peaks at day 13, with a slope of $1.99\times10^{-1}$ day$^{-2}$ from onset.
$R_{t}$ then dissipates more slowly at a rate of  $-3.94\times10^{-3}$ day$^{-2}$.
The infection rate peaks at day 26 in \AI{} (twice the amount of time as in \AS) at a rate of $4.52\times10^{2}$ day$^{-2}$,
and dissipates at a rate of $-5.44\times10^{1}$ day$^{-2}$.
From onset to peak, the mortality rate is similar in both cities: $1.34$ day$^{-2}$ in \AS{} (peak: day 28) and $1.34$ day$^{-2}$ in \AI{} (peak: day 34).

\begin{figure}[ht!]

  \begin{subfigure}{.45\linewidth}
    \centering \small \textsf{\textit{Auto Sprawl}}
  \end{subfigure}
  \begin{subfigure}{.45\linewidth}
    \centering \small \textsf{\textit{Auto Innovative}}
  \end{subfigure}

  \begin{subfigure}{.8\linewidth}
    \includegraphics[width=\textwidth]{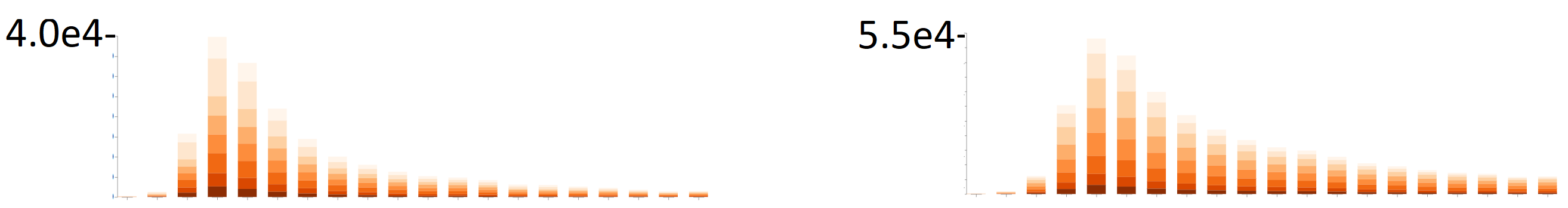}
  \end{subfigure}
  \begin{subfigure}{.17\linewidth}
    \includegraphics[width=\textwidth]{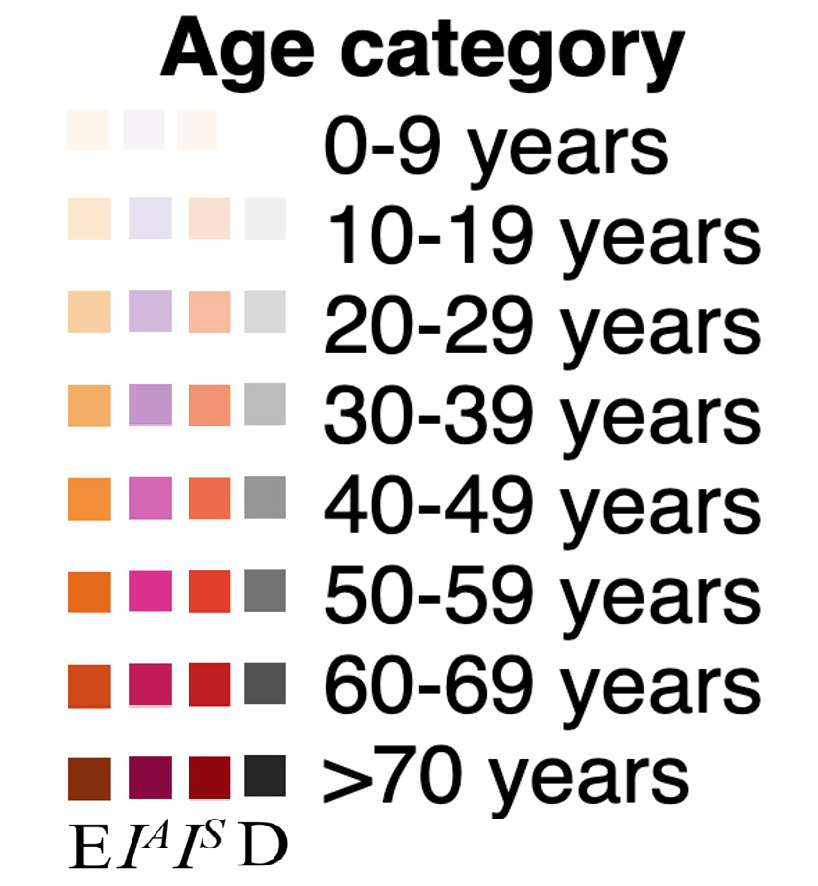}
  \end{subfigure}

  \begin{subfigure}{.8\linewidth}
    \includegraphics[width=\textwidth]{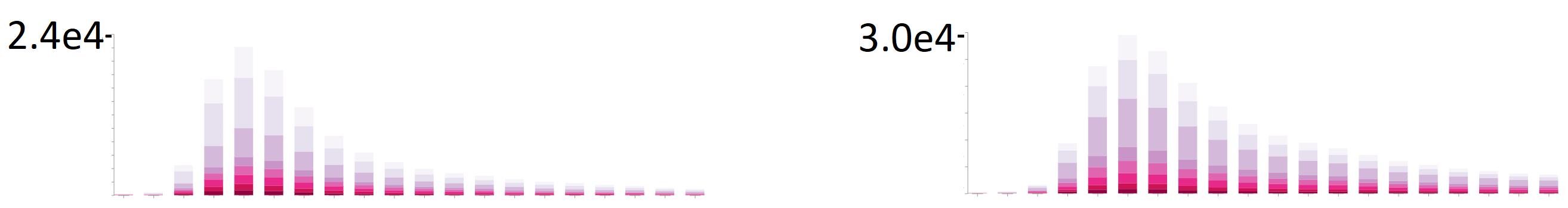}

    \includegraphics[width=\textwidth]{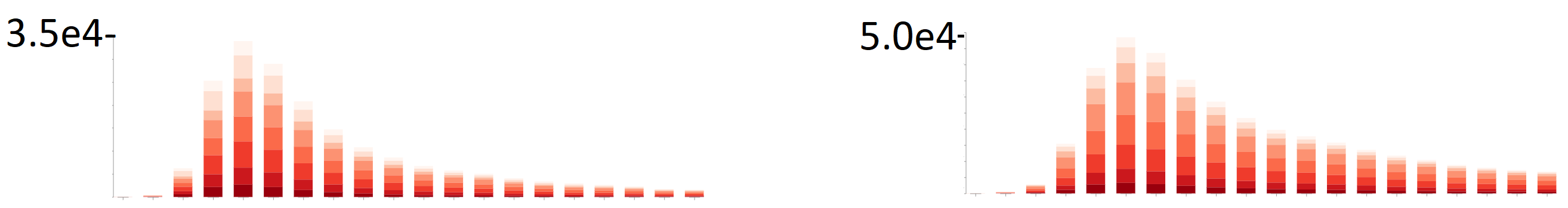}

    \includegraphics[width=13.05cm]{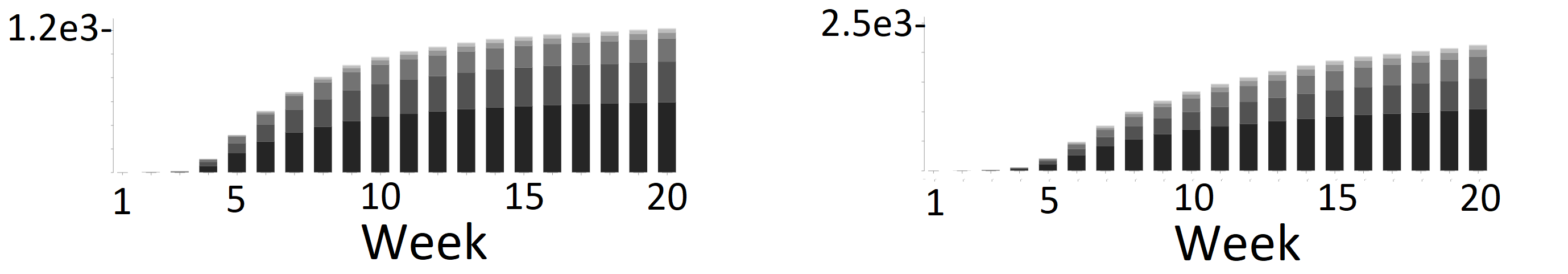}
   \end{subfigure}
  
  
  
   
  \caption{\textbf{Propagation of COVID-19 by age}.
  }
  \label{fig:temporal_age_act}
\end{figure}

Given that we used age-specific probabilities (for symptomaticity and mortality) in our SEIRD model, we can also observe the evolution of COVID-19 by age group in \autoref{fig:temporal_age_act}a-d.
Susceptibilities are distributed as expected, with the oldest age category facing the highest fatalities.
We also consider the evolution of the epidemic by activity (\autoref{fig:temporal_age_act}e-f).
At the onset of the epidemic, \textit{Transit} activity is responsible for most of the transmissions, more so trains than buses.
At peak infection, \textit{Transit} remains responsible for the greatest share of transmissions in \AS, while
\textit{Home} and \textit{Work} have the greatest share in \AI{} (\autoref{fig:temporal_age_act}f).
In the latter stages, we find that the greatest share of infections occurs during \textit{Work}.
\textit{Education} is also significant in \AI{} and \textit{Shopping} in \AS.
The lowest impact activity is \textit{Bus} transit.

\begin{figure}\sf
  \centering
  \begin{subfigure}{\linewidth}
    \caption{}
      {\scriptsize
        \begin{tabular}{cccccc}
            \begin{tikzpicture}
              \draw (0, 0) node[inner sep=0] {\includegraphics[width=2.3cm]{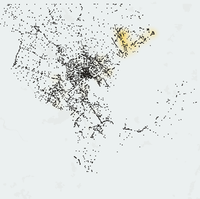}};
              \draw(0, -1) node[inner sep=0] {Day 7};
            \end{tikzpicture}
          &
            \begin{tikzpicture}
              \draw (0, 0) node[inner sep=0] {\includegraphics[width=2.3cm]{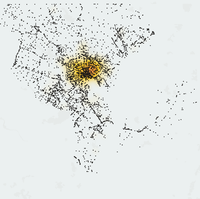}};
              \draw(0, -1) node[inner sep=0] {Day 14};
            \end{tikzpicture}
          &
            \begin{tikzpicture}
              \draw (0, 0) node[inner sep=0] {\includegraphics[width=2.3cm]{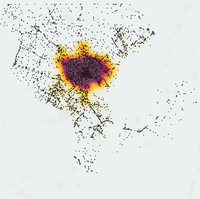}};
              \draw(0, -1) node[inner sep=0] {Day 21};
            \end{tikzpicture}
          &
            \begin{tikzpicture}
              \draw (0, 0) node[inner sep=0] {\includegraphics[width=2.3cm]{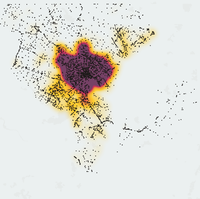}};
              \draw(0, -1) node[inner sep=0] {Day 28};
            \end{tikzpicture}
          &
            \begin{tikzpicture}
              \draw (0, 0) node[inner sep=0] {\includegraphics[width=2.3cm]{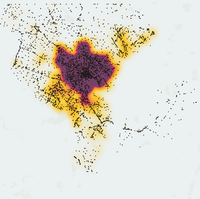}};
              \draw(0, -1) node[inner sep=0] {Day 35};
            \end{tikzpicture}
          &
            \begin{tikzpicture}
              \draw (0, 0) node[inner sep=0] {\includegraphics[width=2.3cm]{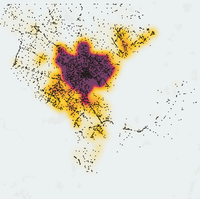}};
              \draw(0, -1) node[inner sep=0] {Day 42};
            \end{tikzpicture}
          \\
            \begin{tikzpicture}
              \draw (0, 0) node[inner sep=0] {\includegraphics[width=2.3cm]{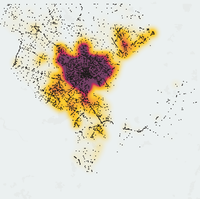}};
              \draw(0, -1) node[inner sep=0] {Day 49};
            \end{tikzpicture}
          &
            \begin{tikzpicture}
              \draw (0, 0) node[inner sep=0] {\includegraphics[width=2.3cm]{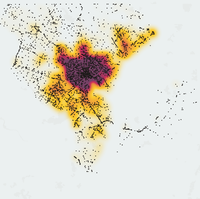}};
              \draw(0, -1) node[inner sep=0] {Day 56};
            \end{tikzpicture}
          &
            \begin{tikzpicture}
              \draw (0, 0) node[inner sep=0] {\includegraphics[width=2.3cm]{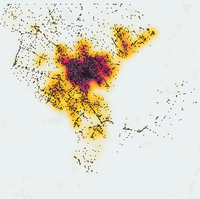}};
              \draw(0, -1) node[inner sep=0] {Day 63};
            \end{tikzpicture}
          &
            \begin{tikzpicture}
              \draw (0, 0) node[inner sep=0] {\includegraphics[width=2.3cm]{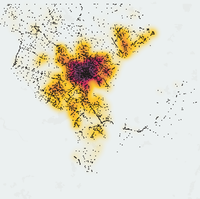}};
              \draw(0, -1) node[inner sep=0] {Day 70};
            \end{tikzpicture}
          &
            \begin{tikzpicture}
              \draw (0, 0) node[inner sep=0] {\includegraphics[width=2.3cm]{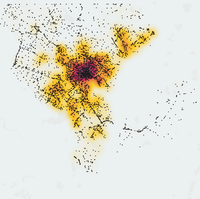}};
              \draw(0, -1) node[inner sep=0] {Day 77};
            \end{tikzpicture}
          &
            \begin{tikzpicture}
              \draw (0, 0) node[inner sep=0] {\includegraphics[width=2.3cm]{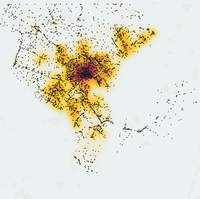}};
              \draw(0, -1) node[inner sep=0] {Day 84};
            \end{tikzpicture}
          \\
            \begin{tikzpicture}
              \draw (0, 0) node[inner sep=0] {\includegraphics[width=2.3cm]{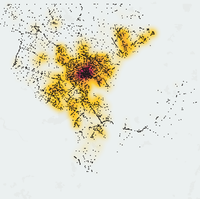}};
              \draw(0, -1) node[inner sep=0] {Day 91};
            \end{tikzpicture}
          &
            \begin{tikzpicture}
              \draw (0, 0) node[inner sep=0] {\includegraphics[width=2.3cm]{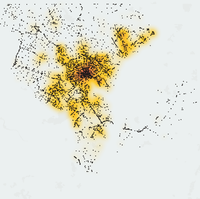}};
              \draw(0, -1) node[inner sep=0] {Day 98};
            \end{tikzpicture}
          &
            \begin{tikzpicture}
              \draw (0, 0) node[inner sep=0] {\includegraphics[width=2.3cm]{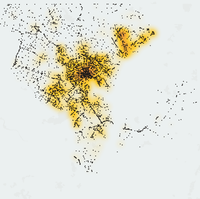}};
              \draw(0, -1) node[inner sep=0] {Day 105};
            \end{tikzpicture}
          &
            \begin{tikzpicture}
              \draw (0, 0) node[inner sep=0] {\includegraphics[width=2.3cm]{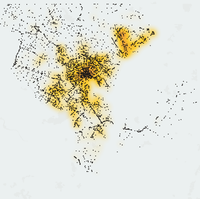}};
              \draw(0, -1) node[inner sep=0] {Day 112};
            \end{tikzpicture}
          &
            \begin{tikzpicture}
              \draw (0, 0) node[inner sep=0] {\includegraphics[width=2.3cm]{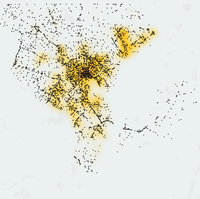}};
              \draw(0, -1) node[inner sep=0] {Day 119};
            \end{tikzpicture}
          &
            \begin{tikzpicture}
              \draw (0, 0) node[inner sep=0] {\includegraphics[width=2.3cm]{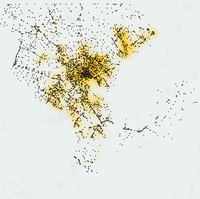}};
              \draw(0, -1) node[inner sep=0] {Day 126};
            \end{tikzpicture} 
          \\
            \begin{tikzpicture}
              \draw (0, 0) node[inner sep=0] {\includegraphics[width=2.3cm]{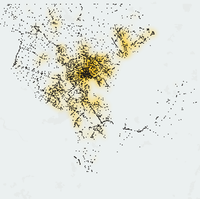}};
              \draw(0, -1) node[inner sep=0] {Day 133};
            \end{tikzpicture}
          &
            \begin{tikzpicture}
              \draw (0, 0) node[inner sep=0] {\includegraphics[width=2.3cm]{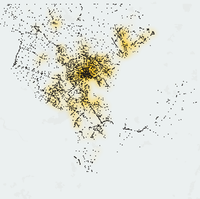}};
              \draw(0, -1) node[inner sep=0] {Day 140};
            \end{tikzpicture}
          &
            \begin{tikzpicture}
              \draw (0, 0) node[inner sep=0] {\includegraphics[width=2.3cm]{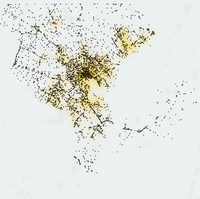}};
              \draw(0, -1) node[inner sep=0] {Day 147};
            \end{tikzpicture}
          &
            \begin{tikzpicture}
              \draw (0, 0) node[inner sep=0] {\includegraphics[width=2.3cm]{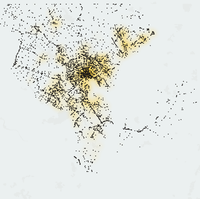}};
              \draw(0, -1) node[inner sep=0] {Day 154};
            \end{tikzpicture}
          &
            \begin{tikzpicture}
              \draw (0, 0) node[inner sep=0] {\includegraphics[width=2.3cm]{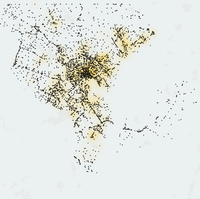}};
              \draw (0, -1) node[inner sep=0] {Day 161};
            \end{tikzpicture}
          &
            \begin{tikzpicture}
              \draw (0, 0) node[inner sep=0] {\includegraphics[width=2.3cm]{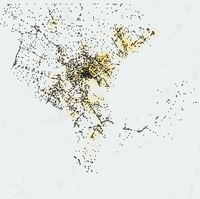}};
              \draw (0, -1) node[inner sep=0] {Day 168};
            \end{tikzpicture} \\
        \end{tabular}
        
      }
    \end{subfigure}
    
  \vspace{2ex}
  
  \begin{subfigure}{\linewidth}
    \caption{}
      {\scriptsize
        \begin{tabular}{cccccc}
            \begin{tikzpicture}
              \draw (0, 0) node[inner sep=0] {\includegraphics[width=2.3cm]{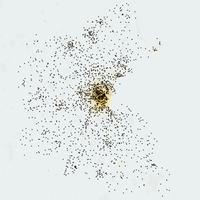}};
              \draw(0, -1) node[inner sep=0] {Day 7};
            \end{tikzpicture}
          &
            \begin{tikzpicture}
              \draw (0, 0) node[inner sep=0] {\includegraphics[width=2.3cm]{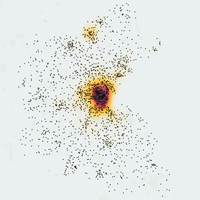}};
              \draw(0, -1) node[inner sep=0] {Day 14};
            \end{tikzpicture}
          &
            \begin{tikzpicture}
              \draw (0, 0) node[inner sep=0] {\includegraphics[width=2.3cm]{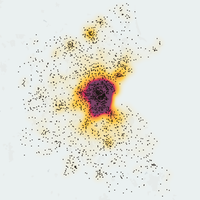}};
              \draw(0, -1) node[inner sep=0] {Day 21};
            \end{tikzpicture}
          &
            \begin{tikzpicture}
              \draw (0, 0) node[inner sep=0] {\includegraphics[width=2.3cm]{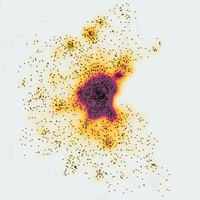}};
              \draw(0, -1) node[inner sep=0] {Day 28};
            \end{tikzpicture}
          &
            \begin{tikzpicture}
              \draw (0, 0) node[inner sep=0] {\includegraphics[width=2.3cm]{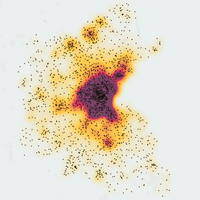}};
              \draw(0, -1) node[inner sep=0] {Day 35};
            \end{tikzpicture}
          &
            \begin{tikzpicture}
              \draw (0, 0) node[inner sep=0] {\includegraphics[width=2.3cm]{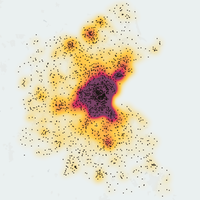}};
              \draw(0, -1) node[inner sep=0] {Day 42};
            \end{tikzpicture}
          \\
            \begin{tikzpicture}
              \draw (0, 0) node[inner sep=0] {\includegraphics[width=2.3cm]{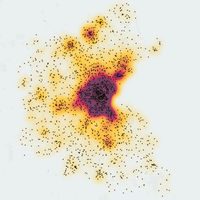}};
              \draw(0, -1) node[inner sep=0] {Day 49};
            \end{tikzpicture}
          &
            \begin{tikzpicture}
              \draw (0, 0) node[inner sep=0] {\includegraphics[width=2.3cm]{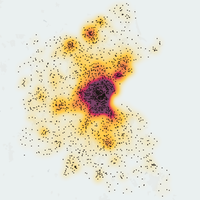}};
              \draw(0, -1) node[inner sep=0] {Day 56};
            \end{tikzpicture}
          &
            \begin{tikzpicture}
              \draw (0, 0) node[inner sep=0] {\includegraphics[width=2.3cm]{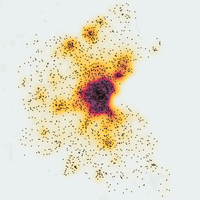}};
              \draw(0, -1) node[inner sep=0] {Day 63};
            \end{tikzpicture}
          &
            \begin{tikzpicture}
              \draw (0, 0) node[inner sep=0] {\includegraphics[width=2.3cm]{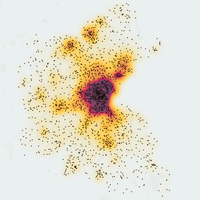}};
              \draw(0, -1) node[inner sep=0] {Day 70};
            \end{tikzpicture}
          &
            \begin{tikzpicture}
              \draw (0, 0) node[inner sep=0] {\includegraphics[width=2.3cm]{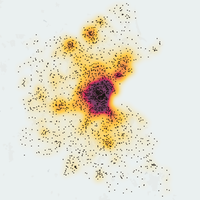}};
              \draw(0, -1) node[inner sep=0] {Day 77};
            \end{tikzpicture}
          &
            \begin{tikzpicture}
              \draw (0, 0) node[inner sep=0] {\includegraphics[width=2.3cm]{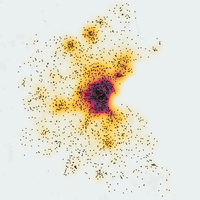}};
              \draw(0, -1) node[inner sep=0] {Day 84};
            \end{tikzpicture}
          \\
            \begin{tikzpicture}
              \draw (0, 0) node[inner sep=0] {\includegraphics[width=2.3cm]{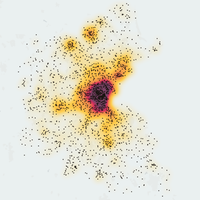}};
              \draw(0, -1) node[inner sep=0] {Day 91};
            \end{tikzpicture}
          &
            \begin{tikzpicture}
              \draw (0, 0) node[inner sep=0] {\includegraphics[width=2.3cm]{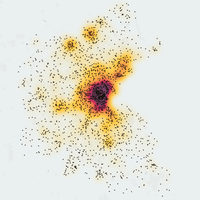}};
              \draw(0, -1) node[inner sep=0] {Day 98};
            \end{tikzpicture}
          &
            \begin{tikzpicture}
              \draw (0, 0) node[inner sep=0] {\includegraphics[width=2.3cm]{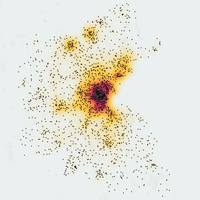}};
              \draw(0, -1) node[inner sep=0] {Day 105};
            \end{tikzpicture}
          &
            \begin{tikzpicture}
              \draw (0, 0) node[inner sep=0] {\includegraphics[width=2.3cm]{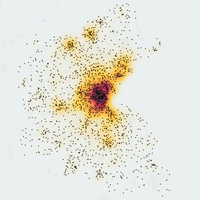}};
              \draw(0, -1) node[inner sep=0] {Day 112};
            \end{tikzpicture}
          &
            \begin{tikzpicture}
              \draw (0, 0) node[inner sep=0] {\includegraphics[width=2.3cm]{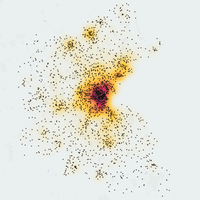}};
              \draw(0, -1) node[inner sep=0] {Day 119};
            \end{tikzpicture}
          &
            \begin{tikzpicture}
              \draw (0, 0) node[inner sep=0] {\includegraphics[width=2.3cm]{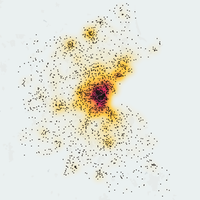}};
              \draw(0, -1) node[inner sep=0] {Day 126};
            \end{tikzpicture} 
          \\
            \begin{tikzpicture}
              \draw (0, 0) node[inner sep=0] {\includegraphics[width=2.3cm]{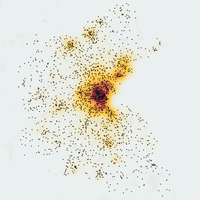}};
              \draw(0, -1) node[inner sep=0] {Day 133};
            \end{tikzpicture}
          &
            \begin{tikzpicture}
              \draw (0, 0) node[inner sep=0] {\includegraphics[width=2.3cm]{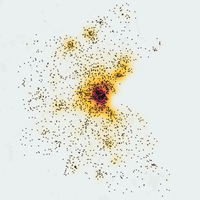}};
              \draw(0, -1) node[inner sep=0] {Day 140};
            \end{tikzpicture}
          &
            \begin{tikzpicture}
              \draw (0, 0) node[inner sep=0] {\includegraphics[width=2.3cm]{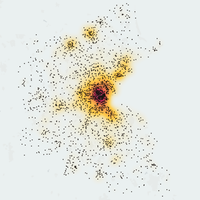}};
              \draw(0, -1) node[inner sep=0] {Day 147};
            \end{tikzpicture}
          &
            \begin{tikzpicture}
              \draw (0, 0) node[inner sep=0] {\includegraphics[width=2.3cm]{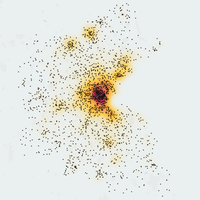}};
              \draw(0, -1) node[inner sep=0] {Day 154};
            \end{tikzpicture}
          &
            \begin{tikzpicture}
              \draw (0, 0) node[inner sep=0] {\includegraphics[width=2.3cm]{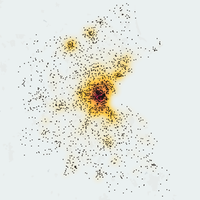}};
              \draw (0, -1) node[inner sep=0] {Day 161};
            \end{tikzpicture}
          &
            \begin{tikzpicture}
              \draw (0, 0) node[inner sep=0] {\includegraphics[width=2.3cm]{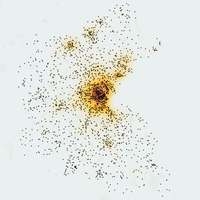}};
              \draw (0, -1) node[inner sep=0] {Day 168};
            \end{tikzpicture} \\
        \end{tabular}
        
      }
    \end{subfigure}
    \caption{\textbf{Spatial evolution of COVID-19.} Heatmap of infected (exposed and infectious) individuals every seven days:
      \textbf{a} \textit{Auto Sprawl}; \textbf{b} \textit{Auto Innovative}.
      As expected, the prevalence is greatest in city centers, regardless of the location of index cases.
      In \textit{Auto Innovative}, the hubs are evident.
    }
    \label{fig:spatial}
  \end{figure}

We observe the spatial evolution of the epidemic in both cities (\autoref{fig:spatial}).
In \AS{}, at the onset of the epidemic, we identify several hubs, mostly located outside of the city center.
These quickly  migrate to the city center by the second week, becoming stronger there, with the peak of the disease being observed in the sixth week.
In \AI, a denser city with greater transit usage, the onset of the epidemic begins in city center, grows rapidly and spreads out to the suburbs.
The peak of epidemic observed between weeks 5 and 7, after which there is a gradual decline of in the numbers throughout the city.
We observe that towards the end of the simulation, the rate of decline is the number of cases is much slower for \AI{} compared to \AS{}.
At the end of our simulation period, \AI{} had around 10 times more active cases compared to \AS{}.


\subsection*{Impact of transit network  on disease propagation}
To test the effect of the physical transit network on the propagation of the epidemic,
we remove the \textit{Transit} activity from the contact network construction in  both \AS{} and \AI{}.
Then, we simulate the same COVID-19 dynamics on these no-transit models.
We compare the SEIRD model outputs and time-dependent reproductive rates (\autoref{fig:temporal}c).
Results indicate that $R_{t}$ peaks on the first day in \AS{} (average value in the first five days is 1.40), dissipating at the rate of $-7.65\times10^{-3}$ day$^{-2}$.
The removal of transit thus effectively dampens the reproduction of the epidemic from the onset in \AS.
In \AI{}, $R_{t}$ peaks at day 25 (compared to day 13 in the full case), but its slope is now $4.29 \times 10^{-2}$ day$^{-2}$, five times smaller than in the full network.
Thus, the force of the epidemic is also lessened in \AI{} by transit removal.

This effect is even more apparent when we consider the infection rates.
In \AS{}, the infection rate peaks at day 86 with $1.49 \times 10^{3}$ per day
(compared to day 18 in the full network case,  with $9.93\times10^4$ infections/day).
Its slope (onset to peak) is $1.16\times10^{1}$ day$^{-2}$ (44 times smaller than in the full case).
Meanwhile, in \AI{}, the infection rate peaks at day 61 with $4.54\times10^{3}$ per day
(compared to day 18 in the full network cases with $1.25\times10^4$ infections/day).
It changes at a rate of $7.12\times10^{1}$ day$^{-2}$ (6.4 times smaller than in the full case).

The force of mortality is also slowed, as daily deaths peak on day 62 in \AS{} (compared to day 28 in the full case) and on day 78 in \AI{} (compared to day 34 in the full case).
These results illustrate the importance of modeling the \textit{Transit} contact network in detail,
and the central role that public transportation plays in spreading the virus.

  

\section*{Discussion}
In order to establish  future epidemic protection programs in large metropolitan areas,
it is necessary to analyze the effectiveness of multiple interventions and epidemic control policies on a large urban scale as well as among different communities and groups. For that purpose, we developed a flexible and dynamic framework, PanCitySim,
which is built on a modified SEIRD model combined with an activity-based mechanistic model, which represents the urban environment
and simulates population mobility and human interactions.
We then introduced a modified SEIRD model incorporating the latest understanding of the propagation of COVID-19.
The epidemiological model enables a computation of infection dynamics of the entire population in the city during their activities (including transit)  at a 5-minute resolution.
We demonstrate the use of PanCitySim and the rich output it provides in two prototype cities representative of most urban areas in the US and Canada (both auto dependent, but one with higher population density and greater share of mass transit).

Crucially, we generate 5-minute activity-based contact networks for entire synthetic populations.
We found the largest average contacts per individual is  90 (in \AS) and 120 (in \AI).
We analyze the degree distributions of these networks, gaining important insights into the mixing of populations across two prototype cities representative of the US and Canada.
Furthermore, we find that the activity contact networks follow well-known distributions describing complex systems.
Contact networks for four activity types were fitted.
Significantly, \textit{Transit} contacts obey the power law ($\hat\alpha \sim 1.4$ up to a maximum of 150 contacts),
which is time-invariant and constant in two distinct city types.
\textit{Shopping} and \textit{Other} follow exponential distributions that are also largely time-invariant.
\textit{Work}, on the other hand, follows a Weibull (stretched exponential) distribution with time-dependent parameters differing by city. While \textit{Work} accounts for the greatest number of contacts per person, particularly during the middle of the day, \textit{Transit}, more than any other activity, accounts for the closest of contacts.  

The inclusion of demographic variables, such as gender and age,
introduces levels of susceptibility in different groups of individuals which is critical for a better understanding of disease propagation.
We observed the dynamics of COVID-19 for 270 days in both cities, and found that even if the index cases begin on the
outskirts of the city, the epidemic rapidly spreads to the city center.
In both cities the epidemic peaks between days 27 and 34 with more than $9\times10^{4}$ infections, dissipating slowly after 150 days if the city is sparse
or after 250 days for a denser typology with more than $1.3\times10^{5}$ exposed or infected individuals.

Further, we found that at the onset of the epidemic it is crucial to restrict mass transit services or focus interventions
(such as enforcing mask-wearing by passengers) on this activity.
Post-peak, however, restrictions should be targeted towards work areas, as well as shopping centers or schools in  less dense car-oriented cities.
Our approach, which is fully mechanistic and highly spatio-temporally resolved, offers insights into the  contact  network  structure and the
importance of having a detailed representation of population mobility. 
With PanCitySim, scenarios can be realistically  modeled and  targeted  to  specific  activities,  ages,  and  employment types.
We can detect the emergence of super-spreading events and show how urban activity patterns affects the spreading of such events.
Further, there is a need to investigate the effects of network topology and  reproduction rates $R_{0}$ and $R_{t}$.
Future work also needs to be done to understand how index case locations affects disease propagation in urban areas.



\section*{Methods}\label{section:methods}
\subsection*{Activity-based simulation model and city data}
The simulation of the prototype cities is performed in SimMobility, an open-source platform for microscopic
demand and mesoscopic supply dynamic traffic assignment modeling \cite{adnan2016simmobility}.  It is calibrated for modeshares, activity
patterns and speeds in the cities we discuss.  The calibration and validation details are available in \cite{oke2020evaluating}.
The activity- and agent-based simulator takes as inputs: land use, demographic and
economic factors, as well as road and transit networks. A discrete choice modeling framework then simulates  daily activity schedules (DAS)
for each individual in a given synthetic population.
Parameters for population (including age, gender, employment, vehicle ownership and household characteristics), land use and road/transit networks are obtained from a real-world archetype city in the typology. 
The DAS is a high level plan, including only important choices, which are  translated into trip chains.
Lower level choices are made during the day when those plans are executed.
During the day, agents are either performing an activity (\textit{Home}, \textit{Work}, \textit{Education}, \textit{Shopping} or \textit{Other})
or executing a trip.
During an activity, agents are stationary at one location and then, at the beginning of each trip, agents
further detail their plan.  Once the detailed plan is made and the start time is reached, the supply simulator
moves the agent accordingly.  Given the path of each traveler, the supply simulator produces the actual
movement trajectory of each heterogeneous vehicle type and pedestrian (passenger) movements, 
which are performed on the network to provide event-driven trajectories for each person.

From the 5-minute person-trajectories, activity-specific contact graphs are constructed.
Transmission events are simulated at this resolution.
Other disease state transitions are simulated at the end of each day.
PanCitySim is a flexible framework that can be readily integrated with other activity-based mobility simulators. The pseudocode for the framework is provided in SI Note 6.

\subsection*{SEIRD model}
We define the following states in our susceptible-exposed-infectious-recovered-deceased (SEIRD) model:
susceptible ($\bm S$), exposed, i.e.\ infected but not contagious ($\bm E$), infectious and symptomatic ($\bm I^{S}$),
infectious but asymptomatic $\bm I^{A}$), recovered ($\bm R$) and deceased ($\bm D$).
We assume that symptomatic cases are automatically quarantined by the end of the day.
    
    
The transitions to each of these states are governed by the following probabilities:
\begin{align}
  \label{eq:40}
  \phi_{n,t} &= 1 - e^{\lt(-\Theta\sum_m q_{m,t}\cdot i_{nm,t}\cdot \tau_{nm,t}\rt)}\\
  \kappa_{n,d} &= 1 - e^{-\fr1{d_L}}\\
  \gamma_{n,d} &= 1 - e^{-\fr1{d_I}} \\
  \mu_{n,d}    &= 1 - e^{-\fr1{d_{D}}}
\end{align}
$\phi$ is the probability of infection, where $\Theta$ is a parameter to be calibrated.
Using the mechanistic framework \cite{muller2020mobility,smieszek2009mechanistic}, $q$ is a measure of viral shedding rate [m$^3$]
and $i$ the contact intensity [$1/$m$^3$]. $\tau$ is the duration of contact. The indices are $m$ (susceptible population), $n$ (infectious) and $t$ (time step).
$\kappa$ is the daily probability of transitioning from exposed $\bm E$ to infectious $\bm I^{\{A,S\}}$ and $d_L$ is the incubation period \cite{prem2020effect}.
Furthermore, $d_L \sim \text{Lognormal} (\mu = 1.62,\sigma^2=0.42)$ \cite{lauer2020incubation}.
The median incubation period is taken as 5 days \cite{lauer2020incubation}.
$\rho$ is the probability an infectious person is symptomatic.
Thus, the transition to state $\bm I^S$ is governed by the probability $\rho\kappa$.
Further, $\alpha$ is the proportion of infectious people who are symptomatic, $\gamma$ is the daily probability that a person recovers ($d_I$ is the duration of infectiousness, also lognormally distributed) and $\mu$ is the mortality rate.
Evidence suggests that $\rho$ and $\mu$ are highly age-dependent \cite{prem2020effect}.
Given that case fatality rates have been shown to differ significantly by age group, we use the values estimated by \cite{verity2020estimates}
to infer the number of days from symptom onset to death $d_{D}$, which is lognormally distributed with age-specific parameters.
Details of the age-specific estimation of $d_{D}$ (along with other SEIRD model explanations) are given in SI Note 1.


\subsection*{Contact intensity}
The transmission probability is explicitly dependent on the separation distance and shedding rate \cite{smieszek2009mechanistic}.
With regard to distance, the contact intensity $i$ has an inverse cube relationship.
For simplicity, we assume the shedding rate ($q_{0}$) remains constant, but estimate the separating distance $d_{n,m}$ at time step $t$ between two persons $n$ and $m$ at a given physical node $V$ based on locations which are assumed to be normally distributed about the respective node center.
We then  calibrate for a fixed contact intensity $i_{0}$, which we then scale by $\frac{1}{d_{nm}^{3}}$ per node and time step.
For agents in transit, we partition the vehicles into squares, randomly assign passengers to each square partition, and estimate the expected distance between randomly selected passenger pairs. When agents are at home, we use the expected distance between inhabitants based on the average square footage of homes in the city.
Thus, the probability of transmission is given by:
\begin{equation}
  \label{eq:20}
  \phi_{n,t} =  1 - \exp\lt\{-\Theta q_0 i_{0} \sum_m  \fr{1}{d_{n,m}^{3}} \cdot \tau_{nm,t} \rt\}
\end{equation}
Further details on the modeling and estimation of inter-person distances in transit vehicles and at activity locations are given in SI Note 2.

\subsection*{Contact network generation}
A contact network is a collection of several contact graphs at different times of day. We generate contact graphs at 5 minute time steps for the entire population. For generation of graphs, we use the outputs from activity-based simulated trajectories from calibrated models of two cities. The contact graph, at any 5 minute time window, is created by a union of three sub graphs- home, activity or transit.
The \textit{Home} graph consists of agents who share a home during the time window.
The activity graph comprises individuals who are performing an activity (\textit{Work}, \textit{Education}, \textit{Shopping} or \textit{Other}).
The \textit{Transit} graph consists of individuals who are traveling in a bus or train or waiting for the same. 
We assume that motorists make no contacts while en route to their destinations, hence they are not modeled in our contact graph.
To facilitate an efficient representation of the contact graphs, we employ a hub-and-spoke transformation, which leverages on the sparseness of the graphs.
Details of the construction of the contact network are found in SI Note 3.

\subsection*{Calibration}
To find $\Theta$, we calibrate in order to achieve a basic reproductive number (average number of secondary cases caused by an infected person in the early stage of the epidemic) of $R_0 = 2.5$ using:
\begin{equation}
  \label{eq:31}
  R_0 = \fr1{X}\sum_m^X\sum_n^S \lt( 1 - e^{-\Theta'\sum_m \tau_{nm}}\rt)
\end{equation}
where $\Theta' = \Theta {q}_{0}{i}_{0}$, and $X$ is the set of index cases while $S$ the set of secondary cases \cite{smieszek2009mechanistic}.
We define $R_0$ as the average number of new infections per initially infected person.
We calibrate $R_0$ over a period of 5 days on the full population, as this is the median duration of incubation \cite{lauer2020incubation}.
Post calibration,  $R_0$ was equal to 2.43 $\pm$ 0.14 in case of \AS{} and 3.19 $\pm$ 0.16 in case of \AI{} using 30 samples for both cases (errors are 95\% confidence intervals).
The five-day average of $R_{t}$ (daily reproductive rate), however was 0.77 in both cities.
Further calibration considerations are noted in SI Note 4.

\subsection*{Validation}
We partially validate the output from PanCitySim.
The partial nature of the validation is due to two facts.
First, the data available for the number of infections are only partially observed.
This is attributed to the overstretching of health facilities, many cases going undetected and the cause of death not being attributed to COVID-19 due to several reasons. Second, as soon as the general population and the government are aware of the epidemic, several interventions are put in place to counter the spread.
Since the testing of such mitigation measures as scenarios was out of the scope of this study, we do not fully match the reported case and mortality numbers.

\begin{figure}\centering
  \begin{subfigure}{.45\linewidth}
    \caption{}
   \includegraphics[width=\textwidth]{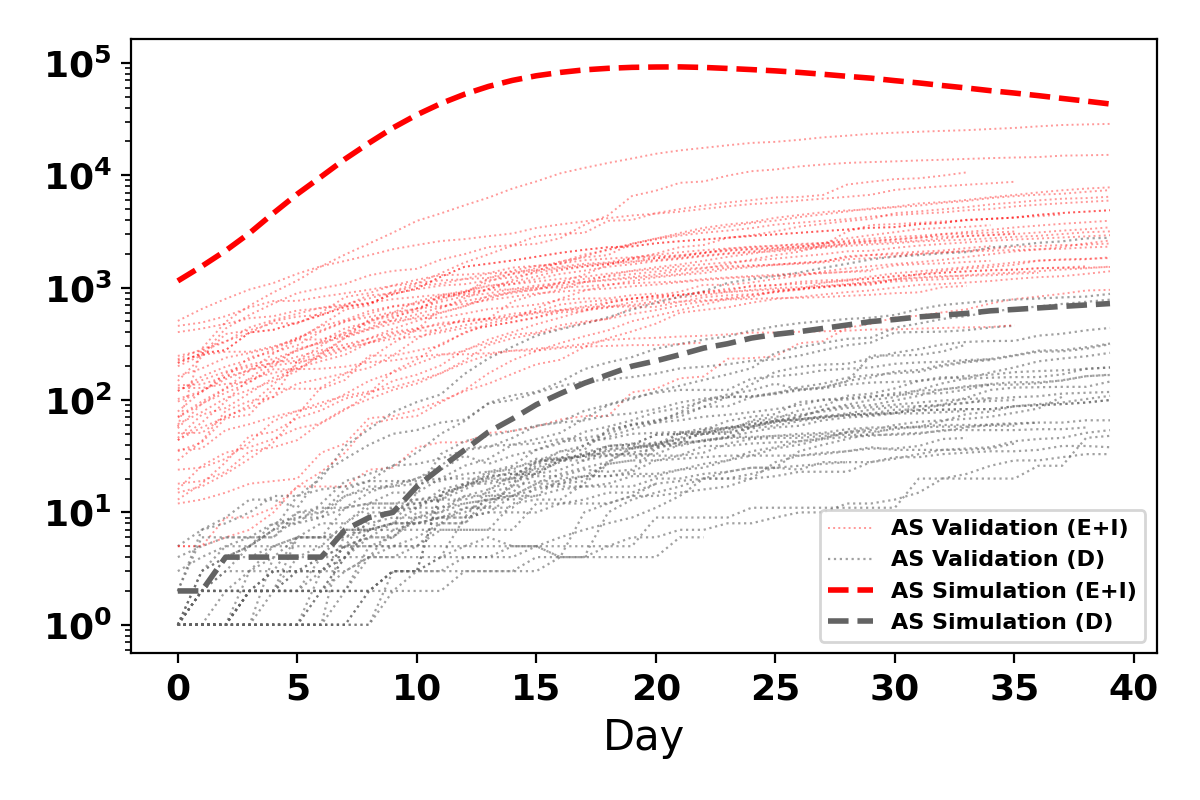}
\end{subfigure}
\begin{subfigure}{.45\linewidth}
  \caption{}
   \includegraphics[width=\textwidth]{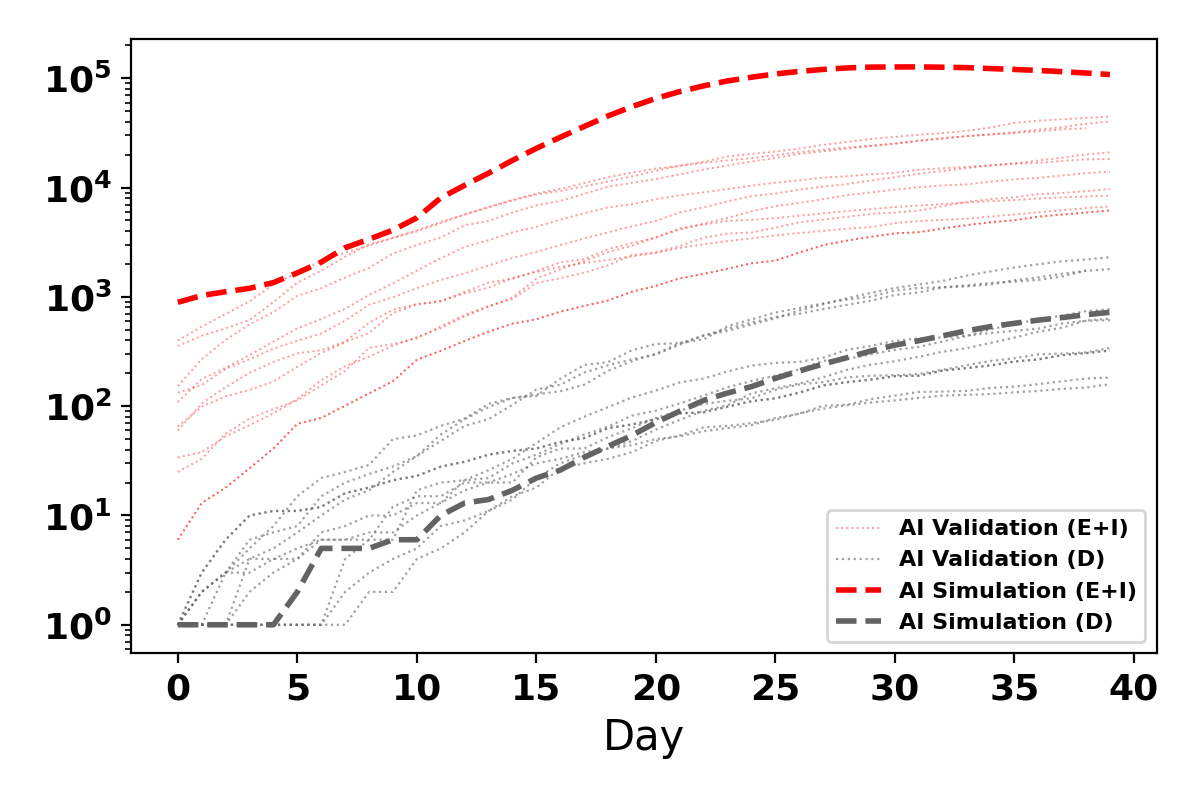} 
\end{subfigure}
\caption{\small We compare simulation results to reported data on cases and mortalities.
  (a) \textit{Auto Sprawl} (simulation) compared to other U.S.\ metropolitan areas in the same typology (e.g.\ Baltimore, Austin, Indianapolis).
  (b)
  \textit{Auto Innovative} (simulation) compared to other U.S.\ metropolitan areas in the same typology (e.g.\ Boston, Chicago, San Francisco)
  Validation start dates were matched with earliest record of the first death.
  Data was obtained from the New York Times repository (\url{https://github.com/nytimes/covid-19-data }).
  \label{fig:validation_data}}

\end{figure}

We compare our prototype city simulation results with data reported from cities in the respective typologies:
\AS{} and \AI{}.
Our findings are summarized in \autoref{fig:validation_data}.
We also note that these contact networks are based on weekday activity models.
However, since the framework is simulator agnostic, we can account for differences in weekend activity patterns by incorporating
relevant travel models.
Finally, to improve the prediction capabilities of this framework, weather patterns should be accounted for in order to reflect appropriate
multimodal impacts on mobility and activity patterns.

\subsection*{Data availability}
The full data set used for this study is publicly available.
Given that the sizes of the data files used in our experiments are large,  we have created a demonstration data set with
30000 individuals chosen randomly from \AS{}, which can be downloaded from  our Github repository for PanCitySim: \url{https://github.com/pancitysim/PanCitySim}.
The links to the file-server hosting the full data set will also be maintained in the same location.

\subsection*{Code availability}
An end-to-end working Python notebook with code and documentation of the epidemiological model is available for public download at
\url{https://github.com/pancitysim/PanCitySim}.
We discuss the computational performance of PanCitySim in SI Note 5, and provide pseudocode for the framework in SI Note 6.




\bibliographystyle{ieeetr}
\bibliography{draft.bib}

\begin{thebibliography}{10}

\bibitem{sahin20202019}
A.~R. Sahin, A.~Erdogan, P.~Mutlu~Agaoglu, Y.~Dineri, A.~Y. Cakirci, M.~E.
  Senel, R.~A. Okyay, and A.~M. Tasdogan, ``2019 {{Novel Coronavirus}}
  ({{COVID}}-19) {{Outbreak}}: {{A Review}} of the {{Current Literature}},''
  {\em EURASIAN JOURNAL OF MEDICINE AND ONCOLOGY}, vol.~4, no.~1, pp.~1--7,
  2020.

\bibitem{mckibbin2020global}
W.~J. McKibbin and R.~Fernando, ``The {{Global Macroeconomic Impacts}} of
  {{COVID}}-19: {{Seven Scenarios}},'' {{SSRN Scholarly Paper}} ID 3547729,
  {Social Science Research Network}, {Rochester, NY}, Mar. 2020.

\bibitem{hackl2019epidemic}
J.~Hackl and T.~Dubernet, ``Epidemic {{Spreading}} in {{Urban Areas Using
  Agent}}-{{Based Transportation Models}},'' {\em Future Internet}, vol.~11,
  p.~92, Apr. 2019.

\bibitem{muller2020mobility}
S.~A. Muller, M.~Balmer, A.~Neumann, and K.~Nagel, ``Mobility traces and
  spreading of {{COVID}}-19,'' {\em medRxiv}, p.~2020.03.27.20045302, Mar.
  2020.

\bibitem{smieszek2009mechanistic}
T.~Smieszek, ``A mechanistic model of infection: Why duration and intensity of
  contacts should be included in models of disease spread,'' {\em Theoretical
  Biology and Medical Modelling}, vol.~6, p.~25, Nov. 2009.

\bibitem{adnan2016simmobility}
M.~Adnan, F.~C. Pereira, C.~L. Azevedo, K.~Basak, M.~Lovric, S.~Raveau, Y.~Zhu,
  J.~Ferreira, C.~Zegras, and M.~{Ben-Akiva}, ``{{SimMobility}}: {{A
  Multi}}-scale {{Integrated Agent}}-{{Based Simulation Platform}},'' in {\em
  Transportation {{Research Board}} 95th {{Annual MeetingTransportation
  Research Board}}}, no.~16-2691, 2016/00/00.

\bibitem{smieszek2011reconstructing}
T.~Smieszek, M.~Balmer, J.~Hattendorf, K.~W. Axhausen, J.~Zinsstag, and R.~W.
  Scholz, ``Reconstructing the 2003/2004 {{H3N2}} influenza epidemic in
  {{Switzerland}} with a spatially explicit, individual-based model,'' {\em BMC
  infectious diseases}, vol.~11, p.~115, May 2011.

\bibitem{prem2020effect}
K.~Prem, Y.~Liu, T.~W. Russell, A.~J. Kucharski, R.~M. Eggo, N.~Davies,
  S.~Flasche, S.~Clifford, C.~A.~B. Pearson, J.~D. Munday, S.~Abbott, H.~Gibbs,
  A.~Rosello, B.~J. Quilty, T.~Jombart, F.~Sun, C.~Diamond, A.~Gimma, K.~van
  Zandvoort, S.~Funk, C.~I. Jarvis, W.~J. Edmunds, N.~I. Bosse, J.~Hellewell,
  M.~Jit, and P.~Klepac, ``The effect of control strategies to reduce social
  mixing on outcomes of the {{COVID}}-19 epidemic in {{Wuhan}}, {{China}}: A
  modelling study,'' {\em The Lancet Public Health}, vol.~0, Mar. 2020.

\bibitem{oke2019novel}
J.~B. Oke, Y.~M. Aboutaleb, A.~Akkinepally, C.~L. Azevedo, Y.~Han, P.~C.
  Zegras, J.~Ferreira, and M.~E. {Ben-Akiva}, ``A novel global urban typology
  framework for sustainable mobility futures,'' {\em Environmental Research
  Letters}, vol.~14, p.~095006, Sept. 2019.

\bibitem{verity2020estimates}
R.~Verity, L.~C. Okell, I.~Dorigatti, P.~Winskill, C.~Whittaker, N.~Imai,
  G.~{Cuomo-Dannenburg}, H.~Thompson, P.~G.~T. Walker, H.~Fu, A.~Dighe, J.~T.
  Griffin, M.~Baguelin, S.~Bhatia, A.~Boonyasiri, A.~Cori, Z.~Cucunub{\'a},
  R.~FitzJohn, K.~Gaythorpe, W.~Green, A.~Hamlet, W.~Hinsley, D.~Laydon,
  G.~{Nedjati-Gilani}, S.~Riley, S.~van Elsland, E.~Volz, H.~Wang, Y.~Wang,
  X.~Xi, C.~A. Donnelly, A.~C. Ghani, and N.~M. Ferguson, ``Estimates of the
  severity of coronavirus disease 2019: A model-based analysis,'' {\em The
  Lancet Infectious Diseases}, vol.~0, Mar. 2020.

\bibitem{lauer2020incubation}
S.~A. Lauer, K.~H. Grantz, Q.~Bi, F.~K. Jones, Q.~Zheng, H.~R. Meredith, A.~S.
  Azman, N.~G. Reich, and J.~Lessler, ``The {{Incubation Period}} of
  {{Coronavirus Disease}} 2019 ({{COVID}}-19) {{From Publicly Reported
  Confirmed Cases}}: {{Estimation}} and {{Application}},'' {\em Annals of
  Internal Medicine}, Mar. 2020.

\bibitem{nahmias-biran2020evaluating}
B.-h. {Nahmias-Biran}, J.~B. Oke, N.~Kumar, C.~Lima~Azevedo, and
  M.~{Ben-Akiva}, ``Evaluating the impacts of shared automated mobility
  on-demand services: An activity-based accessibility approach,'' {\em
  Transportation}, Apr. 2020.

\bibitem{oke2020evaluating}
J.~B. Oke, A.~P. Akkinepally, S.~Chen, Y.~M. Aboutaleb, Y.~Xie, C.~L. Azevedo,
  C.~Zegras, J.~Ferreria, and M.~Ben-Akiva, ``{Evaluating the systemic effects
  of automated mobility-on-demand services via large-scale agent-based
  simulation of auto-dependent prototype cities},'' 2019.

\bibitem{alstott2014powerlaw}
J.~Alstott, E.~Bullmore, and D.~Plenz, ``Powerlaw: {{A Python Package}} for
  {{Analysis}} of {{Heavy}}-{{Tailed Distributions}},'' {\em PLOS ONE}, vol.~9,
  p.~e85777, Jan. 2014.

\bibitem{broido2019scalefree}
A.~D. Broido and A.~Clauset, ``Scale-free networks are rare,'' {\em Nature
  Communications}, vol.~10, p.~1017, Mar. 2019.

\bibitem{kucharski2020early}
A.~J. Kucharski, T.~W. Russell, C.~Diamond, Y.~Liu, J.~Edmunds, S.~Funk, R.~M.
  Eggo, F.~Sun, M.~Jit, J.~D. Munday, N.~Davies, A.~Gimma, K.~van Zandvoort,
  H.~Gibbs, J.~Hellewell, C.~I. Jarvis, S.~Clifford, B.~J. Quilty, N.~I. Bosse,
  S.~Abbott, P.~Klepac, and S.~Flasche, ``Early dynamics of transmission and
  control of {{COVID}}-19: A mathematical modelling study,'' {\em The Lancet
  Infectious Diseases}, vol.~0, Mar. 2020.

\bibitem{lin2020explaining}
G.~Lin, A.~T. Strauss, M.~Pinz, D.~A. Martinez, K.~K. Tseng, E.~Schueller,
  O.~Gatalo, Y.~Yang, S.~A. Levin, E.~Y. Klein, and F.~t. C. M.-H. Program,
  ``Explaining the {{Bomb}}-{{Like Dynamics}} of {{COVID}}-19 with {{Modeling}}
  and the {{Implications}} for {{Policy}},'' {\em medRxiv},
  p.~2020.04.05.20054338, Apr. 2020.

\bibitem{gunzler2020timevarying}
D.~Gunzler and A.~R. Sehgal, ``Time-{{Varying COVID}}-19 {{Reproduction
  Number}} in the {{United States}},'' {\em medRxiv}, p.~2020.04.10.20060863,
  Apr. 2020.

\end{thebibliography}

\section*{Acknowledgments}
The authors acknowledge the support of the University of Massachusetts Amherst, Ariel University and Singapore-ETH Centre in providing resources for carrying out this work. The authors would also like to acknowledge the support of Prof.\ Martin Raubal, ETH Zurich.

\section*{Author contributions}
B. N-B., J.O. and N. K. designed the study and implemented the method.
N. K. implemented the simulations.
J.O. and B. N-B. supported with data acquisition.
B. N-B., J.O. and N. K. analyzed the results and wrote the manuscript.

\section*{Competing interests}
The authors declare no competing interests.

\appendix 

\end{document}
